\documentclass[preprint] {article}
\usepackage{amsmath}
\usepackage{amsfonts}
\usepackage{amssymb}
\usepackage{geometry}
\usepackage{graphicx}
\allowdisplaybreaks
\newcommand {\bp}{\begin{pmatrix}}
\newcommand {\ep}{\end{pmatrix}}
\newcommand{\be}{\begin{equation}} \newcommand{\ee}{\end{equation}}
\newcommand{\bea}{\begin{eqnarray}}\newcommand{\eea}{\end{eqnarray}}

\begin{document}
\title{Hamiltonian formulation of systems with balanced loss-gain
and exactly solvable models }

\author{ Pijush K. Ghosh\footnote {{\bf email:} 
pijushkanti.ghosh@visva-bharati.ac.in} \ and 
Debdeep Sinha\footnote{{\bf email:}  debdeepsinha.rs@visva-bharati.ac.in}} 
\date{Department of Physics, Siksha-Bhavana, \\ 
Visva-Bharati University, \\
Santiniketan, PIN 731 235, India.}
\maketitle

\begin{abstract}

A Hamiltonian formulation of generic many-body systems with balanced loss
and gain is presented. It is shown that a Hamiltonian formulation is
possible only if the balancing of loss and gain terms occur in a pairwise
fashion. It is also shown that with the choice of a suitable co-ordinate,
the Hamiltonian can always be reformulated in the background of a
pseudo-Euclidean metric.  If the equations of motion of  some of the well-known
many-body systems like Calogero models are generalized to include balanced
loss and gain, it appears that the same may not be amenable to a Hamiltonian
formulation. A few exactly solvable systems with balanced loss and gain,
along with a set of integrals of motion is constructed. The examples include a coupled chain of nonlinear
oscillators and a many-particle Calogero-type model with four-body inverse square plus two-body 
pair-wise harmonic interactions. For the case of nonlinear oscillators, stable
solution exists even if the loss and gain parameter has unbounded upper range.
Further, the range of the parameter for which the stable solutions are
obtained is independent of the total number of the oscillators. 
The set of coupled nonlinear equations are solved exactly for the case when the values of all the
constants of motions except the Hamiltonian are equal to zero. Exact, analytical classical solutions 
are presented for all the examples considered. 
\end{abstract}

{\bf keywords:} Hamiltonian formulation, Exactly solvable models, Dissipative system,
Coupled nonlinear oscillators, Calogero-type model

\tableofcontents
\vspace{0.3in}

\section{Introduction}

The Hamiltonian formulation of dissipative systems is a recurring theme
in physics\cite{bat,morse}. One of the earliest approaches to the problem is to
introduce an auxiliary system as a thermal bath that is time-reversed version
of a dissipative harmonic oscillator. The original and the auxiliary oscillators
considered together is described in terms of a Hamiltonian,
when the loss and gain are balanced equally. The rates of dissipation from the
system and absorption by bath are same and the total energy of the system-bath
combination is conserved. Quantization of the Hamiltonian involves various
subtle issues with important results and applications\cite{bopp,feshback,trikochinsky,
dekker,rasetti,rabin,jur}. 

Studies on systems invariant under the combined operation of
parity(${\cal{P}}$) and time-reversal (${\cal{T}}$) symmetry is currently
an active area of research\cite{aa}-\cite{zno} with plethora of experimental results in the
context of superconductivity\cite{super},\cite{super1}, optics\cite{new7}-\cite{zh}, 
microwave cavities\cite{micro}, atomic diffusion\cite{atomic},
nuclear magnetic resonance\cite{reso}, coupled electronic and mechanical oscillators\cite{JC,zh},
coupled whispering galleries\cite{ben,wgm} etc. It is within the context of experiments
on coupled whispering gallery modes, the Hamiltonian of dissipative harmonic
oscillators with balanced loss and gain was reconsidered. The Hamiltonian is
invariant under combined ${\cal{PT}}$-symmetry which the solutions do not obey.
Consequently, the Hamiltonian neither admits classically stable solutions, nor
quantum bound states. However, if  the two oscillators are coupled through
${\cal{PT}}$ symmetric interactions, classically stable solutions as well as
quantum bound states exist within the ranges of the parameters for which 
${\cal{PT}}$ symmetry remains unbroken. An equilibrium is reached for the case
of unbroken ${\cal{PT}}$ symmetry so that the amount of energy transferred to the
bath is reverted back to the system at the equal rate. This mathematical model
explains well the results obtained in Ref. \cite{wgm} and opens up several
new possibilities.

Hamiltonians with balanced loss and gain describe a class of systems which are
intermediate between open and closed systems. In the theory of quantum
dissipation\cite{legg,lang}, the system is coupled to bath consisting of infinitely many harmonic
oscillators so that the energy can not be transferred back to the system.
On the other hand, the sole objective of adding ${\cal{PT}}$ symmetric
interaction to the systems with balanced loss and gain is to achieve an
equilibrium for which the energy is reverted back to the system from the bath
at the same rate as it is deposited to the bath from the system. For the case
of broken ${\cal{PT}}$ symmetry, the equilibrium is lost and the Hamiltonian
is suitable for describing open system.

Dissipation is a natural phenomenon and with the technological advancements,
tailoring systems with balanced loss and gain is a reality. Within this 
background, it is utmost important to study more systems with balanced loss
and gain. It may be recalled at this point that a system of coupled nonlinear
oscillators with balanced loss and gain was investigated in Ref. \cite{khare}
with some interesting results. In absence of any Hamiltonian formulation of
this model, another system of nonlinear oscillators with balanced loss and gain
was analyzed in Ref. \cite{freda} which admits a Hamiltonian formulation.
A chain of linear oscillators with its continuum limit has been studied in
Ref. \cite{ben1}. It has been shown recently that two-body rational Calogero
model with balanced loss and gain terms is exactly solvable admitting 
classically stable solutions as well as  quantum bound states in the
unbroken ${\cal{PT}}$ phase\cite{ds-pkg}. In spite of all of these 
developments, a Hamiltonian
formulation of generic many-particle systems with balanced loss and gain in a
model independent way is lacking. It may be noted that coupled non-linear
Schr$\ddot{o}$dinger equations with balanced loss and gain describe important
physical effects in ${\cal{PT}}$ symmetric theory. The Lagrangian and
Hamiltonian formulation of such systems can be achieved in a straightforward
way within the formulation of non-relativistic field theory. However, the issue
of incorporating balanced loss and gain terms to generic many-particle classical
and quantum mechanical systems has not been addressed in the literature
with its full generality.

The purpose of this article is to present a systematic investigation on
Hamiltonian formulation of generic many-particle classical and quantum
mechanical systems with balanced loss and gain. The form of a generic
many-particle Hamiltonian including gauge potentials is assumed and conditions
are imposed so that the resulting equations of motion contain balanced loss and
gain terms. This severely restricts the allowed ranges of gain-loss
coefficients and types of interaction potentials. The balancing of loss and
gain terms necessarily occurs in a pair-wise fashion, i.e. corresponding to the
co-efficient $\gamma$ of a loss term, there necessarily exists a gain term with
the co-efficient $-\gamma$. The balancing of $\gamma$ by two or more terms is
strictly forbidden. Further, it appears that some of the well known
many-particle systems like Calogero models\cite {calo,sut,ob,poly,pkg1} are not
amenable to Hamiltonian formulation for three or more particles, if they are
generalized to include balanced loss and gain terms. One important aspect of
Calogero model is that each particle interacts with rest of the particles and
it is manifested in the respective equations of motion. If this feature is
relaxed, Hamiltonian formulation of many-body systems governed by
inverse-square plus harmonic interaction with balanced loss and gain terms may
be constructed. In fact, in spite of the imposing restrictions, a Hamiltonian
formulation is possible for a very large class of systems with balanced loss
and gain. The known examples are reproduced within the formulation presented
in this article and some new examples are presented. Finally, it is shown that
the Hamiltonian can always be reformulated in the background of a
pseudo-Euclidean metric through some co-ordinate transformations. A Hamiltonian
formulation for systems with space-dependent balanced loss and gain is also
presented.

The systems with balanced loss and gain terms are investigated from the
viewpoint of exactly solvable models. In particular,
exactly solved systems are constructed for specified forms of potentials.
Two classes of exactly solvable models with balanced loss and gain are
presented which are characterized by invariance of the potential under specific
symmetry transformations: (I) translation and (II) rotation in a
pseudo-Euclidean space endowed with the metric $g_{ij}=(-1)^{i+1} \delta_{ij}$\cite{bla}.
It should be mentioned here that the action or the Hamiltonian of type-I 
systems is not invariant under translation, only the potential respects
this symmetry. Similarly, the Hamiltonian for type-II models is not invariant
under rotation in any plane in the pseudo-Euclidean metric, but, invariant
under rotations in specific planes. The approach taken in constructing exactly
solvable models for $N=2 m, m \in \mathbb{Z^+}$ number of particles is the
following. Apart from the Hamiltonian, $m$ number of integrals of motion
are constructed for type-I as well as type-II models. These integrals of
motion are in involution and imply that the system is at least partially
integrable. The existence of these integrals of motion allows the dynamics
of the original system with $2m$ degrees of freedom to be determined in terms
of an effective system with $m$ degrees of freedom. Consequently, the
exactly solvability of the effective system ensures the same for the original
Hamiltonian. This result is quite general and innumerable numbers of exactly
solvable models with balanced loss and gain can be constructed.

Stable classical solutions are obtained in
terms of Jacobi elliptic functions for particular ranges of parameters for the
type-I models. For the case of a single quartic nonlinear oscillator, stable
solutions are allowed even if the gain-loss parameter $\gamma$ is varied
without any upper bound. It may be recalled at this point that for the case
of linear oscillator\cite{ben}, $\gamma$ can not exceed the value of the
angular frequency of the harmonic oscillator for having stable solutions. The
effect of the non-linearity is also significant for the case of coupled chain of
nonlinear oscillators with uniform gain-loss parameters. Unlike the example
considered in Ref. \cite{ben1}, the region in the parameter-space admitting
stable solutions remains same for any number of particles. Exact classical
solutions are obtained in closed analytical form for several models which are
appropriate generalizations of well-known systems like, two and three particles
coupled nonlinear oscillators, Henon-Heils system, Calogero-type models etc.
The exact solutions of some of the many particle systems with balanced loss and gain 
and are interacting via a four-body potential are obtained.
The exact solutions of these model are generated by exploiting the known solutions
 of Calogero type of models. It is found that the balancing the loss and gain term in
a pair-wise fashion highly restricts the possible form of the four-body interaction.

Exact solutions can be found for several type-II models, including coupled
chain of nonlinear oscillators. However, the solutions are not stable and there
are no region in the parameter-space for which stable solutions are possible.
There are equal number of growing and decaying solutions for even
number of particles. Exact solutions are obtained in terms of Jacobi elliptic
functions multiplied by an exponentially decaying/growing factor. It may be
recalled that a system of  harmonic oscillators with balanced loss and gain
and without any coupling between the two has similar unstable solutions\cite{
bat,morse}. The specified type of interaction allowed for type-II models do
not lead to an equilibrium in respect of energy-transfer between the growing
and the decaying modes and hence, the system is unstable.
 
The plan of the article is the following. The general Hamiltonian formulation
of a generic system with balanced loss and gain is discussed in the next
section. The imposing restrictions on possible forms of Hamiltonian are
discussed in some detail in terms of properties of some matrices appearing
in it. Possible representations of these matrices are discussed in Sec. 2.1.
The condition for a known Hamiltonian to remain a Hamiltonian system, when
it is generalized by including balanced loss and gain is discussed in Sec. 2.2.
The condition is solved for a few cases and some previously studied
systems are reproduced. The condition for Hamiltonian formulation
of rational Calogero model with balanced loss and gain is given in Sec. 2.3
and solved for two particles. In absence of any solution of this condition for
more than two particles, an example of a many-particle system with
inverse-square plus harmonic interaction, which is different from rational
Calogero model, is also presented. Exactly solvable models with stable 
solutions are presented in Sec. 3.1, while models with unstable solutions are
presented in Sec. 3.2. Finally, in Sec. 4, the results are summarized and
implications of the results as well as future directions are discussed. The
results related to a few exactly solvable models are presented in Appendix-A and 
Appendix-B. The Lax-pair formulation of Sutherland 
is considered in Appendix-C and in Appendix-D, $D_N$-type Calogero model 
with four-body interaction is discussed.

\section{Hamiltonian Formulation}

A systematic investigation on the Hamiltonian and Lagrangian formulation of
many-particle systems with balanced loss and gain is presented in this section.
The approach taken in this article is quite general and applicable to
a large class of many-particle systems with balanced loss and gain. 
The Hamiltonian is assumed to be of the form,
\bea
H=\Pi^T M\Pi+V(x_1, x_2, \dots, x_N),
\label{pi}
\eea
where $X=(x_1,x_2, ....x_N)^T$ and $\Pi=(\pi_1,\pi_2, ....\pi_N)^T$ are
$N$ coordinates and their conjugate momenta, respectively. The suffix $T$
denotes transpose, i.e $X^T=Transpose(X)$. The study of constrained systems
is beyond the scope of the present article and hence, the  $N\times N$ real
symmetric matrix $M$ is taken to be non-singular. Any non-standard form of
Hamiltonian is also excluded from the purview of present article.
The generalized momenta $\Pi$ is defined as, 
\bea
\Pi=P+AX,
\label{gen-mom}
\eea
\noindent where $P^T=(p_1, p_2, \dots, p_N)$  is the conjugate momenta and
A is a real $N\times N$ square matrix. Further restrictions on $M$ and $A$
will follow by demanding that $H$ describes a many-particle system with
balanced loss and gain terms. 

The Hamiltonian (\ref{pi}) may be re-expressed as,
\bea
H=P^TMP+X^T A^T MP+P^T MAX+X^TA^TMAX+V(x_1, x_2, \dots, x_N).
\label{Ham}
\eea
and the resulting Eqs. of motion has the following form:
\bea
&& \ddot{X}-2 M R \dot{X}+ 2 M \left ( \frac{\partial V}{\partial X}
\right )=0,\nonumber \\ 
&& R := A - A^T, \ \
\frac{\partial V}{\partial X} \equiv \left (\frac{\partial V}
{\partial x_1}, \frac{\partial V}{\partial x_2},
\dots, \frac{\partial V}{\partial x_N} \right )^T. 
\label{eqm_mat}
\eea
\noindent The matrix $A$ can be decomposed as the sum of a symmetric matrix
$A_s$ and an anti-symmetric matrix $A_a$. The matrix $A_s$ does not contribute
to the Eq. of motion (\ref{eqm_mat}). Further,  $A_s$ can be gauged away from
$\Pi$ in Eq. (\ref{gen-mom}) and hence, from $H$ in the corresponding quantum
theory. In particular,
\bea
 e^{-i X^T A_s X} \Pi e^{i X^T A_s X} \rightarrow
\tilde{\Pi}:=P + A_a X=-i \frac{\partial}{\partial X} + A_a X, \ \
\frac{\partial}{\partial X}:= \left ( \frac{\partial}{\partial x_1},
\frac{\partial}{\partial x_2}, \dots,\frac{\partial}{\partial x_N} \right )^T,
\eea
\noindent where $\hbar=1$. 
Thus, the matrix $A$ can be chosen to be anti-symmetric without loss of any
generality. The choice $A=\frac{R}{2}$ is consistent with the definition of
anti-symmetric matrix $R$. The equation of motion in equation (\ref{eqm_mat})
may also be derived from the Lagrangian,
\be
{\cal{L}} = \frac{1}{4}\dot{X}^T M^{-1} \dot{X} +\frac{1}{4} (X^T R \dot{X}-\dot{X}^T R X) -
V(x_1, x_2, \dots, x_N),
\label{lag}
\ee
\noindent which is invariant under the operation of transposition, i.e.,
${\cal L}^T={\cal L}$, since $R$ is an anti-symmetric matrix by construction.
The parity and time-reversal symmetry may be defined as,
\bea
{\cal{T}}: t \rightarrow - t, \ \ {\cal{P}}: X \rightarrow \tilde{X}= W X,
\eea
\noindent where $W$ is an $N \times N$ orthogonal matrix with determinant $-1$.
The invariance of the Lagrangian ${\cal L}$ and $H$ under ${\cal{PT}}$ implies that
$W$ commutes with $M$, while anti-commutes with $R$. 

A phase-space analysis of Eqs. (\ref{Ham}) and (\ref{eqm_mat}) may be
performed with the assumption that dissipation is due to terms
linear in velocity in the equations of motion. The result shows that only
the diagonal elements of the matrix $MR$ are relevant for determining
whether or not the system is dissipative. 
The following condition is imposed by demanding that the velocity
dependent term in the Eq. of motion for $x_i$ should only contain
$\dot{x}_i$:
\bea
D=M R, \ \ D:= diagonal(\gamma_1,\gamma_2,.....\gamma_N), \ \gamma_i \in \Re.
\label{con1}
\eea
\noindent 
It can be shown that the matrices $M$, $R$ and $D$ anti-commute with
each other:
\bea
\{M,R\}=0,\ \   \{R,D\}=0,\ \   \{M,D\}=0.
\label{anticom}
\eea
\noindent The first relation is derived by using the fact that $M$ and $D$ 
are symmetric matrices, while $R$ is an anti-symmetric matrix. Rest of the
relations follow from the first one and Eq. (\ref{con1}). All the above
relations in Eq. (\ref{anticom}) are true even if $D$ is non-diagonal, but a
symmetric matrix. The anti-commutation relation involving $M$ and $D$ leads
to trace less condition on $D$, 
\be
Tr(D)=0, 
\ee
\noindent implying that the gain and loss are balanced,
i.e $\sum_{i}\gamma_i=0$. It can be further shown that {\it the balancing of
loss and gain terms necessarily occurs in a pairwise fashion for even $N$}. In
particular, if ${\bf v}$ is an eigen-vector of $D$ with non-vanishing
eigen-value $\lambda$, then $M \bf{v}$ is also an eigen-vector of $D$ with
eigenvalue $-\lambda$. Thus, the non-vanishing $\gamma_i$'s are grouped into
$N/2$ independent pairs $\gamma_i+\gamma_j$ such that no two indices are
repeated twice and each such pair is equal to zero. The anti-commutation
relation between $M$ and $D$ also implies that at least one of the $\gamma_i$'s
must be equal to zero for odd N for which $\det(D)=0$:
\be
\det(D) \left [ 1 -(-1)^N \right ]=0.
\ee
\noindent Further, it follows from Eq. (\ref{con1}) that $\det(R)=0$ for
$\det(D)=0$, since $M$ is non-singular. Both $R$ and $D$ can be chosen to
be non-singular for even $N$ and the following relations hold:
\be
Tr(M)=0, \ \ \{R^{-1}, D \}=0.
\ee
\noindent Further, for even $N$, the minimum and maximum numbers of independent
non-vanishing off-diagonal elements of $M_{ij}$ are $N/2$ and $N^2/4$,
respectively. The same result for odd $N$ may be obtained by substituting
$N \rightarrow N-1$ in the respective expression for even $N$. For odd
$N$, the element $M_{NN} \neq 0$. 

A few comments are in order before proceeding further:\\
(1)The anti-commutation relation of $M$ with $R$ or $D$ for $N= 2m, m \in 
\mathbb{Z^+}$ implies that corresponding to each of its $m$ positive
eigenvalues $\lambda$, there exists an eigenvalue $-\lambda$. The 
diagonal matrix $M_d=\hat{O} M \hat{O}^T$ corresponding to $M$ has the
expression $M_d=diagonal(\lambda_1, -\lambda_1, \lambda_2,-\lambda_2,$ 
$\dots, \lambda_m, -\lambda_m ) $, where $\hat{O}$ is an orthogonal
matrix and a particular ordering of eigenvalues is assumed. Any other ordering
obtained by permuting the eigenvalues in the expression of $M_d$ is also
acceptable. With the introduction of the anti-symmetric matrix $\tilde{R}=
\hat{O} R \hat{O}^T$ and a set of new co-ordinates,
\be
\tilde{X} = \hat{O} X, \ \tilde{P} = \hat{O} P, \
\tilde{\Pi} = \hat{O} \Pi=\tilde{P} + \frac{1}{2} \tilde{R} \tilde{X},
\label{newcoor}
\ee
\noindent the Hamiltonian $H$ in Eq. (\ref{pi}) may be re-interpreted as
defined in the background of an indefinite metric $M_d$:
\bea
H & = &\tilde{\Pi}^T M_d \tilde{\Pi} + V(\tilde{x}_1, \tilde{x}_2, \dots, 
\tilde{x}_N)\nonumber \\
& = & \tilde{P}^T M_d \tilde{P} + \tilde{X}^T Q \tilde{P} +
\frac{1}{4} \tilde{X}^T \left ( M_d \tilde{R}^2 \right ) \tilde{X}
+ V(\tilde{x}_1, \tilde{x}_2, \dots, \tilde{x}_N),
\label{newhamil}
\eea
\noindent where $Q:= M_d \tilde{R}$ and $\{M_d, \tilde{R}\}=0$. 
The matrix $Q$ is symmetric with all of its diagonal
elements equal to zero, while $M_d \tilde{R}^2$ is a symmetric matrix.
It may be recalled that a similar indefinite metric appears in the Hamiltonian
formulation of Pais-Uhlenbeck oscillator\cite{puo}.

(2) Systems with space-dependent dissipation coefficient exhibit interesting
physical effect and has various applications. Two common examples from this
class of systems are van der pol oscillator\cite{vander} and van der Pol-Duffing
oscillator\cite{dvander}. The coefficient acts as a gain term in some region of space,
while it is dissipative in some other region of space. Such a system may be
generalized to include balanced loss and gain terms by considering a
time-reversed version of it as a thermal bath with a Hamiltonian formulation
for the system-bath combination. The loss term in some region of space for
the system is equally balanced by a gain term for the bath in the same region
of space and vice verse. The Hamiltonian is still given by ({\ref{pi}) with
$\Pi$ in Eq. (\ref{gen-mom}) replaced by,
\be
\Pi= P + A F, \ \ F \equiv ( F_1(x_1, \dots, x_N), F_2(x_1, \dots, x_N),
\dots, F_N(x_1, \dots, x_N))^T
\ee
\noindent where $F$ is a $N$ component column vector with each component
depending on the dependent variables. A similar analysis as above shows
that the equation of motion reads,
\be
\ddot{X}-2 Q D \dot{X}+ 2 M \left ( \frac{\partial V}{\partial X}
\right )=0, \ \ Q \equiv\sum_{i=1}^N \frac{\partial F_i}{\partial x_i}.
\label{van}
\ee
\noindent System with space-dependent gain-loss term is obtained for
suitable choices of the function $F_i$. Note that when $F_i=\frac{x_i}{N}$, the system
reduces to the form of system having constant loss and gain coefficient.

\def\badmash{Note that the matrix $D$ is expressed as a product of a symmetric
matrix $M$ and and an anti-symmetric matrix $R$. It is a well known
fact\cite{LR} that if a matrix $B$ can be expressed as a product of a
symmetric and an anti-symmetric matrix, then it is similar to $-B$. It
immediately follows that,  
\bea
D= - S D S^{-1}, \implies
Tr(D)=0,\ \ \ \ \det(D)[1-(-1)^N]=0,
\label{con2}
\eea
\noindent where $S$ is a non-singular matrix. The first condition implies
that the gain and loss are balanced, i.e $\sum_{i}\gamma_i=0$.
The last condition implies that at least one of the  $\gamma_i$'s must
be equal to zero for odd N for which $\det[D]=0$.  Further, it follows
from Eq. (\ref{con1}) that $\det[D]=0$ also implies $\det[R]=0$, since
$M$ is non-singular. There are further restrictions on the possible values
of $\gamma_i$'s and elements of $M$ due to Eq.  (\ref{con1}). It can be
shown that for even $N$ and $\det[R] \neq 0$ (or equivalently 
$\gamma_i \neq 0 \ \forall \ i$),
\bea
\{M,R\}=0,\ \   \{R,D\}=0,\ \   \{M,D\}=0 .    
\label{anticom}
\eea
\noindent The first equation above implies $Tr[M]=0$. Under the same condition,
the last equation $M_{ij}(\gamma_i+\gamma_j)=0$ shows
$M_{ii} = 0 \ \forall \ i$ and the balancing
of loss and gain terms  necessarily occurs in a pairwise fashion, i,e.
for even $N$ the non-vanishing $\gamma_i$'s are grouped into $N/2$
independent pairs $\gamma_i+\gamma_j$ such that no two indices are repeated
twice and each such pair is equal to zero.}

\subsection{Representation of Matrices}

The condition (\ref{con1}) can be realized for $N=2$ with the following choice: 
\be
M= \sigma_x, \ \ R=-i \gamma \sigma_y, \ \ D= \gamma \sigma_z,
\label{choice0}
\ee
\noindent where $\gamma_1=-\gamma_2\equiv \gamma$ and
$\sigma_x,\sigma_y,\sigma_z$ are Pauli spin matrices. 
Such a realization can be easily generalized to any even $N=2 m, m \in 
\mathbb{Z}^+$: 
\be
M= \kappa \otimes \sigma_x, \ \ D= \eta \otimes \sigma_z, \ \ 
R=-i (\kappa^{-1} \eta ) \otimes \sigma_y, 
\label{choice1}
\ee
\noindent where the commutator of $m \times m$ diagonal matrix
$\eta := diagonal(\gamma_1, \gamma_3, \dots, \gamma_{m} )$ 
and  the $m \times m$ invertible real symmetric matrix $\kappa$ must vanish,
i.e.,
\be
\left [\kappa, \eta \right ] = 0.
\label{commu}
\ee
\noindent The representations of $M$ and $D$ in Eq. (\ref{choice1}) are chosen,
while that of $R$ is determined by using the Eq. (\ref{con1}), i.e.
$R=M^{-1} D$. The pair-wise balancing of loss and gain terms impose the
conditions, $\gamma_{2i} = -\gamma_{2i-1}, i=1, 2, \dots m$. The constraint
(\ref{commu}) can be solved in various ways. For example, $\eta= 
\gamma I_m$ solves Eq. (\ref{commu}) for any $\kappa$, where $I_m$ is
$m \times m$ identity matrix. This choice corresponds to uniform gain and
loss parameter $\gamma$, i.e. $\gamma_{2i-1} = \gamma \ \forall \ i$. On the
other hand, $\kappa=I_m$ solves Eq. (\ref{commu}) for any $\eta$
with  $\gamma_{2i} = -\gamma_{2i-1}, i=1, 2, \dots m$. This corresponds to
a system with non-uniform gain and loss parameters.

 It may be recalled that
the Hamiltonian $H$ in Eq. (\ref{pi}) can be re-interpreted as
defined in the background of an indefinite metric $M_d$. 
For the representation (\ref{choice1}) with  $\eta= 
\gamma I_m$ and  $\kappa=I_m$, the diagonal matrix
$M_d=\hat{O}M\hat{O}^T$ and the orthogonal matrix $\hat{O}$ has the following expressions: 
\bea
M_d=I_m\otimes\sigma_z,\ \  \  \ \hat{O}=\frac{1}{\sqrt{2}}\left[I_m\otimes\left(\sigma_x+\sigma_z\right)\right].
\label{MO}
\eea
The matrix $\hat{O}$ generates the new coordinates and momenta as defined by Eq. (\ref{newcoor}). The new Hamiltonian is given by
Eq. (\ref{newhamil}) with  
\bea
Q=M_d\tilde{R}=\gamma I_m\otimes \sigma_x,\ \ \ \ \ M_d\tilde{R}^2=-\gamma^2 I_m\otimes \sigma_z,
\label{transcoor}
\eea
and has the following form
\bea
H & = & \tilde{P}^T[I_m\otimes\sigma_z]  \tilde{P} + \gamma \tilde{X}^T[I_m\otimes \sigma_x] \tilde{P} -
\frac{\gamma^2 }{4} \tilde{X}^T \left[I_m\otimes \sigma_z \right]\tilde{X}
+ V(\tilde{x}_1, \tilde{x}_2, \dots, \tilde{x}_N).
\label{newhamil1}
\eea
The first and third terms of Eq. (\ref{newhamil1}) are quadratic in momenta and coordinate respectively and
give the length in a pseudo Euclidean space. The second term is a sum of product of coordinates and momenta
and bears the characteristic of balanced loss and gain system.

 It may be noted that 
the representation $(\ref{choice1})$  is not unique and there are several other
possibilities. One such example is presented below:
\bea
&& [M]_{ij}=\delta_{i,j+1} + \delta_{i,j-1}, \ \
[D]_{ij}=(-1)^{i+1} \gamma \delta_{ij}, \ \ 
[R]_{ij}=(-1)^{j+1} [M^{-1}]_{ij},\nonumber \\
&& [M^{-1}]_{ij} = -[M^{-1}]_{ji}=
\frac{1}{4} (-1)^{\frac{j-i-1}{2}} \left [ 1 -(-1)^{j-i} \right ] \left [ 1 -
(-1)^i \right ], \ j > i.
\label{choice2}
\eea
The diagonal matrix $M_d=\hat{O}M\hat{O}^T$ corresponding to the representation (\ref{choice2}), gives rise to
a pseudo Euclidean metric where the form of the matrix $\hat{O}$, generating the new coordinates and momenta
 as defined by Eq. (\ref{newcoor}), is given by:
\bea
\hat{O}_{ij}=\sqrt{\frac{2}{N+1}}\sin\left({\frac{ij \pi}{N+1}}\right),
\eea 
and the form of $M_d$ becomes:
\bea
M_d=diagonal (2\cos{\frac{\pi}{N+1}}, 2\cos{\frac{2\pi}{N+1}},.....,2\cos{\frac{N\pi}{N+1}}).
\eea 
It may be noted that the $i$-th and the $N+1-i$-th eigenvalues of $M_d$ are equal but
opposite in sign. 
The new Hamiltonian is given by Eq. (\ref{newhamil})
with  
\bea
Q=M_d\tilde{R}=\gamma \delta_{i,N+1-i},\ \ \ \ \ M_d\tilde{R}^2=-\gamma^2 M_d^{-1}.
\eea
The Hamiltonian in the pseudo Euclidean space corresponding to
the representations ($\ref{choice1}$) and ($\ref{choice2}$) are not equivalent in the sense that the
determinant of $M_d$ is different in two cases. However, these two  pseudo Euclidean metric
can be made equivalent by simultaneous implication of the following rearrangements:

\noindent i) A proper scale transformation for the representation (\ref{choice2}) of conjugate coordinates and momenta
of the following form:
\bea
&&(\tilde{p}_k,\tilde{p}_{N+1-k})\rightarrow\sqrt{2\cos{\frac{k\pi}{N+1}}}(\tilde{p}_k,\tilde{p}_{N+1-k}),\  \
(\tilde{x}_k,\tilde{x}_{N+1-k})\rightarrow \frac{1}{\sqrt{2\cos{\frac{k\pi}{N+1}}}}(\tilde{x}_k,\tilde{x}_{N+1-k}),\nonumber\\
  &&\hspace{5.5cm} k=1,2,....,m.
\eea
\noindent ii) In case of representation (\ref{choice1}), a particular ordering of the eigenvalues of $M_d$ is assumed. 
If this ordering is made such that the $i$-th and the $N+1-i$-th eigenvalue become equal and opposite in sign.

\noindent Thus, for both the representations (\ref{choice1}) and  (\ref{choice2}), a similarity transformation  to a  pseudo Euclidean 
space followed by a proper scaling of coordinates and momenta will give rise to the same form of the Hamiltonian and hence the same
equations of motion. However, in original coordinates  the representations 
(\ref{choice1}) and  (\ref{choice2}) describe different systems having different form of potentials. The solutions
of these different kind of systems can be found by exploiting the solutions obtained in  a  pseudo Euclidean 
space.

Either of the above representations of matrices for even $N=2m$
can be suitably generalized for odd $N=2m+1$. The $2m+1 \times 2m+1$ 
dimensional matrices $M, R, D$ are constructed by embedding the
respective $2m \times 2m$ dimensional representation in it and choosing the
extra matrix elements as follows:
\bea
&& M_{2m+1,i}=M_{i,2m+1}=0, i=1, 2, \dots 2m, \ \ M_{2m+1,2m+1} \neq 
0,\nonumber \\ 
&& R_{i,2m+1}=R_{2m+1,i}=0, D_{i,2m+1}=D_{2m+1,i}=0, i=1, 2, \dots 2m+1.
\label{choice-odd}
\eea
\noindent The equation governing the dynamics of $x_{2m+1}$ does not contain
any gain or loss term. However, it may be coupled to other degrees of freedom
through appropriate choices of the interaction potential.

\subsection{Examples}

The Eq. of motion (\ref{eqm_mat}) can be re-written as, 
\be
\ddot{X}- 2 D\dot{X}+ 2 M \left ( \frac{\partial V}
{\partial X} \right )=0,
\label{Eom}
\ee
\noindent and equivalently in component form:
\be
\ddot{x}_i - 2 D_{ii} \dot{x}_i + 2 \sum_{k=1}^N M_{ik} 
\frac{\partial V}{\partial x_k}=0, \ i=1, 2, \dots, N.
\label{Eom-c}
\ee
\noindent Unlike in the standard Hamiltonian formulation, the Eq. of motion
for $x_i$ does not contain the term $\frac{\partial V}{\partial x_i}$ for even
$N$ due to the trace less condition on $M$. For odd $N$, only one diagonal
element of $M$ may be chosen to be non-zero. Consequently, if $M_{pp} \neq 0$,
the Eq. of motion for $x_p$ contains the term, 
$\frac{\partial V}{\partial x_p}$. If a system governed by the Hamiltonian
$\tilde{H}=P^T P + \tilde{V}$ is generalized to include balanced loss and gain
terms such that a Hamiltonian formulation in the form of $H$ is possible and
the Eqs. of motion resulting from these two Hamiltonians are identical in the
limit of vanishing loss and gain terms, then
the following condition must hold:
\be
\frac{\partial {V} }{\partial x_i}= \sum_{k=1}^N M^{-1}_{ik}
\frac{\partial \tilde{V}}{\partial x_k} \ i=1, 2, \dots, N.  
\label{kathin-condi}
\ee
\noindent This constitutes a set of coupled first-order differential equations.
These equations are in general nonlinear, except for the case for which the
potential $\tilde{V}$ is a quadratic form of the co-ordinates. The set of
equations can be solved for quadratic $\tilde{V}$. In particular,
\be
\tilde{V} = X^T M G X  \implies  V=X^T G X,
\ee
\noindent where $G$ is an arbitrary symmetric matrix.
The system of coupled ${\cal{PT}}$-symmetric oscillators with 
uniform loss and gain co-efficients and interaction parameters of 
Ref. \cite{ben1} may be obtained by choosing the potential $G$ as,
\be
G =\omega^2 M^{-1} + \epsilon I_N,
\ee
\noindent and using the representation of $M$ given in Eq. (\ref{choice2}),
where $I_N$ is $N \times N$ identity matrix. Similarly, the example of system
with non-uniform parameters may also be reproduced with following  choices of
$M, D, R$:
\be
[M]_{ij} = \delta_{j,N+1-i}, \ D=\delta_{ij} \gamma_i, \
\gamma_{N+1-i}= -\gamma_i, \ R=M^{-1} D=MD.
\ee
\noindent The matrix $G$ for this case is an anti-tridiagonal matrix with
the elements,
\be
G_{ij} = \omega_i^2 \delta_{j,N+1-i} + 
\epsilon_i \left ( \delta_{j,N-i} + \delta_{j,N+2-i} \right ). 
\ee
The formulation may be used to construct Hamiltonian system
corresponding to coupled chain of nonlinear oscillators with balanced loss
and gain terms. For example, a Hamiltonian system for two coupled nonlinear
oscillators may be constructed by considering the representations of $M, R, D$
as in Eq. (\ref{choice0}) and choosing the potential,
\be
V=a_0 x_1 x_2 + a_1 (x_1^2+x_2^2) + a_2 ( x_1 x_2^3 + x_2 x_1^3) + 
a_3 x_1^2 x_2^2 + a_4 (x_1^4 +x_2^4), \ a_i \in \Re.
\ee
\noindent This system is ${\cal{PT}}$-symmetric and reduces to the example
studied in Ref. \cite{freda} for $a_3=0=a_4$. It is also evident that a 
Hamiltonian formulation of the system of nonlinear oscillators considered in
Ref. \cite{khare} is forbidden. There are various possibilities
for generalizing this system to $N$ coupled chain of nonlinear oscillators. A
straightforward generalization would be to consider the potential,
\be
V=\sum_{i=1}^{N-1} \left [ a_0 x_i x_{i+1} + a_2 ( x_i x_{i+1}^3 + x_{i+1} 
x_i^3) + a_3 x_i^2 x_{i+1}^2 \right ] + \sum_{i=1}^N \left [
a_1 x_i^2 + + a_4 x_i^4 \right ], \ a_i \in \Re.
\ee
\noindent and the representations in Eqs. ({\ref{choice2}) and 
({\ref{choice-odd}). The parameters appearing in $V$ are chosen in such a way
that the $H$ is ${\cal{PT}}$-symmetric.\\

System with space-dependent dissipative coefficient can be obtained from Eq. (\ref{van}).
The example of this kind of systems are Van der pol oscillator\cite{vander}, in case of which the dissipative
term has the coefficient $(x^2-1)$, and duffing Van der pol oscillator\cite{dvander}.
Oscillator systems having space dependent dissipative coefficient can easily be
generated for the appropriate form of $F$ in Eq. (\ref{van}). For example,
in case of two coupled oscillators with space dependent balanced loss and gain terms,
the form of $F$ may be considered as
$F_i=\frac{1}{2}x_i-\frac{1}{3}x_i^3$ which gives $Q=[1-(x_1^2+x_2^2)]$. 
The various type of coupling between 
the two oscillators can be generated by choosing appropriate form of the potential $V$.
In this case Eq. (\ref{van}) takes the following form:

\bea
\ddot{x}_1-2\gamma[1-(x_1^2+x_2^2)]\dot{x}_1+\frac{\partial V}{\partial x_2}&=&0,\\
\ddot{x}_2+2\gamma[1-(x_1^2+x_2^2)]\dot{x}_2+\frac{\partial V}{\partial x_1}&=&0,
\eea
where  $M, R, D$ are as in Eq. (\ref{choice0}). It may be noted that inside the circle 
$x_1^2+x_2^2=1$, the first oscillator has a gain and the second oscillator has an
equal amount of loss and out side the circle the situation is reversed.

\subsection{Rational Calogero Model(RCM)}

Given the importance of RCM and its relation to various diverse
areas of physics, it is desirable to find its generalization involving
balanced loss and gain terms. The potential $\tilde{V}_{RCM}$ for the
RCM has the following form:
\bea
\tilde{V}_{RCM} = \frac{\omega^2}{2} \sum_{i=1}^N x_i^2 +
\sum_{\substack{i,j=1 \\ i < j}}^N \frac{g}{(x_i-x_j)^2}.
\eea
\noindent An important feature of RCM is that each particle interacts
with rest of the particles through inverse-square interaction.
The potential $V_{RCM}$ is determined through 
a set of $N$ coupled first-order nonlinear partial differential equations:
\be
\frac{\partial V_{RCM}}{\partial x_i} = \sum_{k=1}^N M^{-1}_{ik} 
\left [ \omega^2 x_k - \sum_{\substack{l=1 \\ l\neq k}}^N
\frac{2 g}{(x_k-x_l)^3} \right ].
\label{diff-calo}
\ee
\noindent For $N=2$, $V_{RCM}$ can be determined easily:
\be
V_{RCM} = \omega^2 x_1 x_2 - \frac{g}{(x_1-x_2)^2},
\label{2rcm}
\ee 
\noindent where the representation of $M$ is given by (\ref{choice0}). The
classical as well as the quantum version of this two-body rational Calogero
model has been studied in detail in Ref. \cite{ds-pkg}. Finding
a solution for $N >2$ and $g \neq0$ is a highly nontrivial problem and it
seems that a consistent solution of Eq. (\ref{diff-calo}) may not exist.
Many-body systems with balanced loss and gain terms may be 
constructed, where particles interact through harmonic plus inverse-square
interaction. However, in the limit of vanishing loss/gain parameters,
the equations of motion differ from that of RCM. These systems may be
identified as Calogero-type models with balanced loss and gain terms.
Such a system may be obtained easily by substituting $\tilde{V}_{RCM}$
for  $V$ in Eqs. (\ref{pi},\ref{Ham},\ref{Eom},\ref{Eom-c}). The resulting
Eqs. of motion with the representation of $M$ as in  Eq. (\ref{choice1}) read:
\bea
&& \ddot{x}_1 - 2 \gamma \dot{x}_1 + 2 \omega^2 x_{2}
-\sum_{\substack{k=1\\ k \neq 2}}^N \frac{4g}{(x_{2}-x_k)^3}=0,\nonumber \\
&& \ddot{x}_i - 2 \gamma (-1)^{i+1} \dot{x}_i + 2 \omega^2
\left ( x_{i-1} + x_{i+1} \right ) - 
\sum_{\substack{k=1\\ k \neq i-1}}^N \frac{4g}{(x_{i-1}-x_k)^3} 
-\sum_{\substack{k=1\\ k \neq i+1}}^N \frac{4g}{(x_{i+1}-x_k)^3}=0,\nonumber\\
&&2 \leq i \leq N-1, \nonumber \\
&& \ddot{x}_N - 2 \gamma (-1)^{N+1} \dot{x}_N + 2 \omega^2 x_{N-1} - 
\sum_{\substack{k=1\\ k \neq N-1}}^N \frac{4g}{(x_{N-1}-x_k)^3}=0. 
\label{rcm-type}
\eea
\noindent The Eqs. of motion of classical RCM is not reproduced in the limit
of vanishing $\gamma$. It may be noted that in Eq. (\ref{Eom}), the column
matrix $\frac{\partial V}{\partial X}$ is multiplied by the off diagonal
matrix $M$ which hinder $N$ particle generalization of RCM with balanced loss
and gain.  In the case of RCM, the dynamics governing the $i$'th
particle contains interaction terms where the $i$'th particle interacts through
pair-wise harmonic plus inverse-square interaction with all other particles.
However, in the present case, the dynamics governing the $i$'th particle
contains interaction terms where its nearest neighbors interact with all
other particles including it through pair-wise harmonic plus inverse-square
interaction. In the limit of vanishing $g$, the coupled chain of linear
oscillators considered in Ref. \cite{ben1} is reproduced. Thus, the Hamiltonian
which gives rise to Eqs. (\ref{rcm-type}) is an appropriate generalization of
the system considered in Ref. \cite{ben1} with inverse-square interaction. 
This constitutes a new class of models which deserves further investigations.
The exact solvability and/or integrability of the system is not apparent and
beyond the scope of the investigations of the present article.

The most general integrable potential for Calogero-Moser-Sutherland(CMS) system is the
Weierstrass elliptic potential of the following form $V(x)={\cal P}(x,2\omega_1,2\omega_2)$.
 This is a single valued doubly periodic function of a single complex variable $x$ with 
$\omega_1$ and $\omega_2$ are being the two half periods. In the limit $\omega_1 
\rightarrow \infty, \omega_2 \rightarrow \infty$, $V(x)\rightarrow x^{-2}$ and we get
the rational CMS model\cite{calo}. The trigonometric or hyperbolic CMS model arises
when either $\omega_1$ or $\omega_2\rightarrow \infty$. Below we discuss the case of 
trigonometric and hyperbolic potentials with balanced loss and gain.

The Sutherland model was originally motivated to extract thermodynamics from
 the model having only the rational interaction term. The Calogero model with only the 
rational potential is not a bound system and one could put the system in a finite 
periodic box, such that the particles interact through all the infinitely many periodic
 images of themselves. In this case the two-body
potential becomes a periodic one. The Sutherland model can easily be generalized to
incorporate the balanced loss and gain term for two particles system. For three
or more particles, the Hamiltonian formulation is possible in the presence of balanced
loss and gain term only if we relax the pair-wise interaction among the particles.
The Sutherland model with trigonometric interaction is given by the potential:

\bea
\tilde{V}_{S} = \sum_{\substack{i,j=1 \\ i < j}}^N \frac{a^2}{\sin^2{a(x_i-x_j)}}.
\eea
The potential $V_S$ can be determined from Eq. (\ref{kathin-condi}):

\be
\frac{\partial V_{S}}{\partial x_i} = \sum_{k=1}^N M^{-1}_{ik} 
\left [\sum_{\substack{l=1 \\ k > l}}^N \frac{4a^3\cos{a(x_k-x_l)}}{\sin^2{a(x_k-x_l)}}-
\sum_{\substack{l=1 \\ k < l}}^N \frac{4a^3\cos{a(x_k-x_l)}}{\sin^2{a(x_k-x_l)}}.
\right ].
\label{diff-calo1}
\ee
This is a set of $N$ coupled first order nonlinear partial differential equations.
 For $N=2$, $V_{S}$ can be determined easily:
\be
V_{S} = - \frac{a^2}{\sin^2{a(x_i-x_j)}},
\label{2rcm1}
\ee 
where the representation of $M$ is given by (\ref{choice0}). For two particles case, 
the vanishing loss and gain parameter will produce the original model. The
solutions of this model with balancing loss and gain term allow
stable solutions. This model will be considered in detail in
Ref. \cite{pkg-ds}. For three or more particles
the solutions of Eq. (\ref{diff-calo1}) is highly nontrivial and it seems that a 
consistent solution may not exist.  A Hamiltonian formulation for many particle system 
having balanced loss and gain and are interacting through a trigonometric potential may be
constructed. However, in this case the vanishing loss and gain parameter will not produce the 
original model. This kind of model may be obtained by replacing $\tilde{V}_S$ by $V_S$ in 
constructing the Eqs of motion. The resulting Eqs. of motion are given by:
\bea
&& \ddot{x}_1 - 2 \gamma \dot{x}_1+\sum_{\substack{l=1 \\ 2 > l}}^N \frac{8a^3\cos{a(x_2-x_l)}}{\sin^2{a(x_2-x_l)}}-
\sum_{\substack{l=1 \\ 2 < l}}^N \frac{8a^3\cos{a(x_2-x_l)}}{\sin^2{a(x_2-x_l)}}=0,\nonumber \\
&& \ddot{x}_i - 2 \gamma (-1)^{i+1} \dot{x}_i
+\sum_{\substack{l=1 \\ i-1 > l}}^N \frac{8a^3\cos{a(x_{i-1}-x_l)}}{\sin^2{a(x_{i-1}-x_l)}}-
\sum_{\substack{l=1 \\ i-1 < l}}^N \frac{8a^3\cos{a(x_{i-1}-x_l)}}{\sin^2{a(x_{i-1}-x_l)}},\nonumber\\
&&+\sum_{\substack{l=1 \\ i+1 > l}}^N \frac{8a^3\cos{a(x_{i+1}-x_l)}}{\sin^2{a(x_{i+1}-x_l)}}-
\sum_{\substack{l=1 \\ i+1 < l}}^N \frac{8a^3\cos{a(x_{i+1}-x_l)}}{\sin^2{a(x_{i+1}-x_l)}}=0\nonumber\\
&&2 \leq i \leq N-1, \nonumber \\
&& \ddot{x}_N - 2 \gamma (-1)^{N+1} \dot{x}_N 
+\sum_{\substack{l=1 \\ N-1 > l}}^N \frac{8a^3\cos{a(x_{N-1}-x_l)}}{\sin^2{a(x_{N-1}-x_l)}}\nonumber\\
&&-\sum_{\substack{l=1 \\ N-1 < l}}^N \frac{8a^3\cos{a(x_{N-1}-x_l)}}{\sin^2{a(x_{N-1}-x_l)}}=0
\label{rcm-type1}
\eea
 where the representation of $M$ is given by (\ref{choice1}). The above model describes
a system where the dynamics of the $i$-th particle involves interacting term where its
nearest neighbor interact with all other particles including it. This constitutes a new class
of model and needs further investigations. A similar construction for hyperbolic
potential is possible in the presence of balanced loss and gain term if we replace $a$ by $ia$ in 
all above calculations. 

The many particle interaction of Calogero model is governed by
root system of finite reflection groups. The type of Calogero model, we consider in this
work belongs to A-type root system. However, Calogero-Moser system is integrable for
other classical root systems (B,C,D)\cite{ob} as well for exceptional and non-crystallographic
 root systems\cite{aj}. For two particles case, various root systems can be modified
to incorporate the loss and gain terms. Detail analysis regarding these models
will be considered in \cite{pkg-ds}. However, all these models are not amenable to 
Hamiltonian formulation for three or more particles in the sense that the systems are not 
reduced to the original many-particle model in the limit of vanishing loss/gain parameter.

\section{Exactly Solvable Models}

The Hamiltonian is a constant of motion. The invariance of the Lagrangian
${\cal{L}}$ under different symmetry transformations may lead to other first
integrals of the system. Invariance under space-translation is ruled out even
for translationally invariant potential $V$. This is due to the presence of
terms linear in $X$ and $X^T$, i.e. the second and the third terms in Eq.
(\ref{lag}). Similarly, invariance under rotation is also ruled out even
for rotationally invariant potentials. Nevertheless, integrals of motion
may be constructed for specific choices of $M$, $D$, $R$ and $V$. A few
such exactly solvable many-particle systems with $N= 2m, m \in \mathbf{Z^+}$
are presented in this section.

%The matrices $M$, $D$ and $R$ are chosen as,
%\be
%M= I_m \otimes \sigma_x, \ \ D= \gamma I_m \otimes \sigma_z, \ \ 
%R=-i \gamma I_m \otimes \sigma_y. 
%\label{mat-mat}
%\ee
 For the representation (\ref{choice1}) with  $\eta= 
\gamma I_m$ and  $\kappa=I_m$, a co-ordinate transformation of the form (\ref{newcoor})
by the matrix $\hat{O}$ as given by Eq. (\ref{MO})
with $\tilde{X}=(z_i^+,z_i^-)^T, i=1,2,......,m$ and $ X=(x_{2i-1},z_{2i})^T, i=1,2,......,m$, generates
the following set of new co-ordinates:
\be
z_i^{-}= \frac{1}{\sqrt{2}} \left ( x_{2i-1}-x_{2i} \right ),
\ \ z_i^{+}=\frac{1}{\sqrt{2}} \left ( x_{2i-1}+x_{2i} \right ), \ i=1, 2,\dots m
\label{trans}
\ee

\noindent Writing Eq. (\ref{newhamil1}) in the component form, the
Hamiltonian is obtained in the background of a pseudo Euclidean metric in the following form:
\be
H=  \sum_{i=1}^m \left [( P_{z_i^+}^2 - P_{z_i^-}^2) +
\gamma \left ( z_i^+ P_{z_i^-} + z_i^- P_{z_i^+} \right ) - 
\frac{\gamma^2}{4} \left \{ (z_i^+)^2 - (z_i^-)^2 \right \} \right ] +
V(z_i^+,z_i^-)
\label{ham-z}
\ee
\noindent The conjugate momenta have the following expressions:
\be
P_{z_i^+} = \frac{1}{2}\left(\dot{z}_i^+ -\gamma z_i^{-}\right), \ \
P_{z_i^-} = -\frac{1}{2}\left(\dot{z}_i^- - \gamma z_i^{+}\right).
\ee
\noindent The Eqs. of motion can be derived from the Hamilton's equations of
motion as,
\be
\ddot{z}_i^{+} -2 \gamma \dot{z}_i^{-} + 
 2\frac{\partial V}{\partial z_i^+}=0,\ \
\ddot{z}_i^{-} - 2 \gamma \dot{z}_i^{+} - 
2 \frac{\partial V}{\partial z_i^-}=0,\ \ i=1, 2, \dots m.
\label{z-eqn}
\ee
\noindent In the subsequent discussions $2V$ is replaced by $V$. 
The Eqs. (\ref{z-eqn}) is a set of $2m$ coupled differential equations. 

The parity (${\cal{P}}$) transformation is defined as,
\bea
{\cal{P}}: && X \rightarrow - \left ( I_{m} \otimes \sigma_x \right ) X, 
P \rightarrow  - \left ( I_{m} \otimes \sigma_x \right ) P, \nonumber \\
&& z_i^+ \rightarrow - z_i^+, \ \ z_i^- \rightarrow z_i^-, \
P_{z_i^+} \rightarrow - P_{z_i^+}, \ \ P_{z_i^-} \rightarrow P_{z_i^-}.
\label{parity-old}
\eea
\noindent For odd $N=2m+1$, the above relations are supplemented with the
transformations for $x_{2m+1}$ and $p_{2m+1}$, $x_{2m+1} \rightarrow x_{2m+1},
p_{2m+1} \rightarrow p_{2m+1} $.
The time-reversal transformation ${\cal{T}}$ is defined as,
\be
{\cal{T}}: X \rightarrow X, \ P \rightarrow - P, \implies
z_i^{+} \rightarrow z_i^{+},z_i^{-} \rightarrow z_i^{-} \ \
P_{z_i^{+}} \rightarrow - P_{z_i^{+}}, \ P_{z_i^{-}} \rightarrow - P_{z_i^{-}}.
\ee
\noindent 
\noindent For odd number of particles $N=2m+1$, the above relations are
supplemented with the relations: $ x_{2m+1} \rightarrow x_{2m+1}, \
p_{2m+1} \rightarrow - p_{2m+1}$. The Hamiltonian is invariant under the
combined operation of ${\cal{PT}}$ provided the real potential $V$ satisfies
the condition:
\be
V(z_i^+,z_i^-) =V(-z_i^+, z_i^-).
\label{condi-pot}
\ee 
\noindent It may be noted that the parity transformation defined by Eq.
(\ref{parity-old}) is not unique. Consequently, the condition on 
$V(z_i^-,z_i^+)$ to be ${\cal{PT}}$ symmetric also varies depending on the
choice of ${\cal{P}}$. For example, the parity transformation ${\cal{P}}_1$
\be
{\cal{P}}_1: z_i^+ \rightarrow z_i^+, \ \ z_i^- \rightarrow - z_i^-, \
P_{z_i^+} \rightarrow P_{z_i^+}, \ \ P_{z_i^-} \rightarrow - P_{z_i^-},
\ee
\noindent keeps $H$ in Eq. (\ref{ham-z}) ${\cal{P}}_1{\cal{T}}$ invariant
provided $V(z_i^+,z_i^-) =V(z_i^+, -z_i^-)$.
Exactly solvable models can be found for several specific choices
of $V$ satisfying the above constraints. A few examples with ${\cal{PT}}$
invariance and the condition (\ref{condi-pot}) are presented below.

\subsection{Translationally invariant potential}

The set of $2m$ equations (\ref{z-eqn}) can be decoupled into $m$ coupled
equations for the choices of $V\equiv V(z_i^-)$ or $V\equiv V(z_i^+)$.
The potential is translationally invariant for $V\equiv V(z_i^-)$. 
In particular, the potential remains invariant under the transformations
$x_{2i-1} \rightarrow x_{2i-1} + \eta_i, x_{2i}\rightarrow x_{2i} +
\eta_i$, where $\eta_i$'s are $m$ independent parameters. The form of the
potential is special in the sense that it allows $m$ independent parameters
$\eta_i$ instead of a single one. The Hamiltonian is translationally
invariant. The potential is 
chosen to be independent of $z_i^+$ co-ordinates and hence, $V$ as well as
$H$ is ${\cal{PT}}$ symmetric.  With the choice of
this translationally invariant potential, the first set of Eqs. of
(\ref{z-eqn}) can be solved as,
\be 
\dot{z}_i^{+} = 2 \gamma {z}_i^{-} + \Pi_i, \ \ \Pi_i \in \Re \ \forall i,
\ee
\noindent which when substituted in the second set of Eqs. result in 
the following set of $m$ coupled differential equations:
\be
\ddot{z}_i^{-} - 4 \gamma^2 {z}_i^{-} - 
 \frac{\partial V}{\partial z_i^-}= 2 \gamma \Pi_i,\ \ i=1, 2, \dots m.
\label{z-reduce}
\ee
\noindent where $\Pi_i$'s are $m$ integration constants to be fixed by imposing
initial conditions. The co-ordinates $z_i^+$ are determined as,
\be
z_i^+(t) = 2 \gamma \int z_i^-(t) dt + \Pi_i t + C_i, \ \ C_i \in \Re \ \forall
i,
\label{eqnzi+}
\ee
\noindent where $C_i$'s are integration constants. The inhomogeneous term on
the right hand side of Eq. (\ref{z-reduce}) poses problem for finding exact
solutions for a large class of $V$ and can be absorbed through a constant
shift of the co-ordinates,
\be
z_i^- \rightarrow z_i= z_i^- + \frac{\Pi_i}{2 \gamma},
\ee
\noindent which reduces Eqs. (\ref{z-reduce}) and (\ref{eqnzi+}) to the
following forms:
\bea
&& \ddot{z}_i - 4 \gamma^2 {z}_i -  \frac{\partial V(z_i)}{\partial z_i}= 0,\
\nonumber \\
&& z_i^+(t) = 2 \gamma \int z_i(t) dt + C_i, \ \ i=1, 2, \dots m.
\label{homo-eqn}
\eea
\noindent However, an unpleasant outcome of this transformation is that the
potential $V$, in general, becomes dependent on the  dissipation parameter
$\gamma$. In order to
avoid such undesirable feature of the potential, the inhomogeneous part on the
right hand side of Eq. (\ref{z-reduce}) is taken to be zero by appropriate
choice of initial conditions. In particular, the initial conditions on
$\dot{z}_i^+(0)$ and $z_i^-(0)$:
\be
\dot{z}_i^+(0) = 2 \gamma z_i^-(0), \ \forall \ i,
\ee
\noindent gives the values of $m$ integration constants $\Pi_i=0$.
Initial conditions on $z_i^+(0)$ and $\dot{z}_i^-(0)$ may be chosen
independently and are to be specified later. It may be noted that
$z_i^-=z_i$ for $\Pi_i=0$ and Eqs. (\ref{homo-eqn}) are considered
for further discussions.

The $\Pi_i$'s are $m$ integrals of motion:
\be
\Pi_i= 2P_{z_i^+} -\gamma z_i^-, \ \{\Pi_i,\Pi_j\}_{PB}=0, \ \ \{H, \Pi_i\}_{PB}=0,
\ee
\noindent where $\{,\}_{PB}$ denotes the Poisson bracket. The Hamiltonian along
with $\Pi_i$'s constitute $m+1$ integrals of motion, implying that the system
is at least partially integrable. It may be noted that the existence of $m$
integrals of motion is a consequence of the invariance of the Hamiltonian
under translations with $m$ independent parameters $\eta_i$. If the potential
is chosen as, $V \equiv V(z_i^+)$, then $H$ is invariant under translations
$z_i^- \rightarrow z_i^- + \eta_i$. The corresponding conserved quantities
$\Pi_i^-= 2P_{z_i^-} + \gamma z_i^+$  are in involution and constitute $m$
integrals of motion implying partial integrability of the system. 
The form of the potential for both the cases discussed above needs to be
explicitly specified in order to explore admissible exactly solvable models.
The dynamical equations satisfied by $z_i^-$ and $z_i^+$ for
$V \equiv V(z_i^+)$ and $V \equiv V(z_i^-)$ are related to each through the
duality relation:
\be
z_i^- \leftrightarrow z_i^+, \ \ V(z_i^-) \leftrightarrow -V(z_i^+), \ \
\Pi_i \leftrightarrow - \Pi_i^-.
\ee
\noindent In particular, equation satisfied by $z_i^-$ and $z_i^+$ for
$V \equiv V(z_i^+)$ may be obtained from Eqs. (\ref{z-reduce}) and
(\ref{eqnzi+}) by using the above transformation and without any change for
$t, C_i$. Further investigations
on exactly solvable models will be restricted to the case $V \equiv V(z_i^-)$
in this article. The duality relation may be used to obtain results for
$V \equiv V(z_i^+)$. There are plenty of choices of $V$ for which  system
governed by Eq.  (\ref{homo-eqn}) is exactly solvable. 

\subsubsection{Quartic Oscillators} 

One of the simplest choices 
for $m=1, N=2$ is to consider the case of cubic nonlinear oscillator:
\bea
&& V(z_1) =-2 \omega_0^2 z_1^2 - \frac{\alpha}{4} z_1^4, \ \omega, \alpha \in 
\Re,\nonumber \\
&& \ddot{z}_1 + \omega^2 {z}_1 + \alpha z_1^3, \  
\omega^2 \equiv 4 ( \omega_0^2-\gamma^2 ).
\label{cubic-o}
\eea
\noindent There are three different parametric regions for which exact
solutions can be obtained in terms of Jacobi Elliptic functions: 

(i) $\omega^2 > 0, \alpha >0$: The gain-loss parameter $\gamma$ is restricted
to lie in the range, $-\omega_0 < \gamma < \omega_0$. \\
\bea
&& z_1(t)= A \ cn (\Omega t,k),\ \
z_1^+(t) = \frac{2 \gamma }{\Omega} \frac{\cos^{-1} \{ dn(\Omega t,k) \}
sn(\Omega t,k)}{\sqrt{1-dn^2(\Omega t,k)}}\nonumber \\
&& \Omega=\sqrt{\omega^2 + \alpha A^2}, \
k^2=\frac{\alpha A^2}{2 \Omega^2},
\label{solex1}
\eea
\noindent Non-singular stable solutions are obtained for $0< k < 1$.\\
(ii) $\omega^2 > 0, \alpha < 0$: The gain-loss parameter $\gamma$ is restricted
to lie in the range, $-\omega_0 < \gamma < \omega_0$.\\
\bea
&& z_1(t) =  A \ sn(\Omega t,k),\
z_1^+(t) = \frac{2 \gamma }{\sqrt{k} \Omega} log \left [ dn(\Omega t,k)
-\sqrt{k} cn(\Omega t,k) \right ] \nonumber \\
&& \Omega=(\omega^2 -\frac{{\mid \alpha \mid} A^2}{2})^{\frac{1}{2}}, \
k^2 = \frac{{\mid \alpha \mid} A^2}{2 \Omega^2 }, \
0 \leq A \leq \sqrt{\frac{\omega^2}{{\mid \alpha \mid}}}  
\label{solex2}
\eea
\noindent The restriction for having non-singular solution is $0 < k <1$.

(iii)  $\omega^2 < 0, \alpha > 0$: Unlike the previous two cases, the angular
frequency is restricted to lie in the range, $-\gamma < \omega_0 < \gamma$.
The solutions are unstable for the case of linear oscillators with balanced
loss and gain\cite{ben}, whenever the gain-loss parameter $\gamma$ exceeds the
value of angular frequency. However, the effect of the nonlinear term is to
allow stable solutions even when $\gamma > \omega_0$. The stable solutions
exists, even if the confining harmonic term is absent, i.e., $\omega_0=0$.
There is no restriction on the possible range of $\gamma$ for this particular
choice of potential. In particular, there are
two different types of solutions depending on the ranges of amplitude $A$ of
the solution of $z_1(t)$, one of which is stable and bounded. The unstable
solution has the form: 
\bea
&& z_1(t) =  A dn(\Omega t,k),\ \ 
z_1^+(t) = \frac{2 \gamma }{\Omega} am(t,k) \nonumber \\
&& \Omega=(\frac{{\alpha } A^2}{2}), \
k^2 = \frac{\alpha A^2-{\mid \omega^2 \mid}}{2 \Omega^2 }, \
\sqrt{\frac{{\mid \omega^2 \mid}}{\alpha}} \leq A \leq 
\sqrt{\frac{2 {\mid \omega^2 \mid}}{\alpha}},
\label{solex3}
\eea
\noindent where $z_1^+(t)$ is unbounded within the range $0 < k < 1$. The
second solution is obtained for $\sqrt{\frac{2 {\mid \omega^2 \mid}}{\alpha}}
\leq A < \infty $ and is similar to Eq. (\ref{solex1}) except for the
expressions for $\Omega$ and $k$. The non-singular, stable solution is obtained
as, 
\bea
&& z_1(t)= A \ cn (\Omega t,k),\ \
z_1^+(t) = \frac{2 \gamma }{\Omega} \frac{\cos^{-1} \{ dn(\Omega t,k) \}
sn(\Omega t,k)}{\sqrt{1-dn^2(\Omega t,k)}}\nonumber \\
\nonumber \\
&& \Omega= \sqrt{-{\mid \omega^2 \mid} + \alpha A^2}, \
k^2=\frac{\alpha A^2}{2 \Omega^2}. 
\eea
\noindent The effect of the non-linear interaction is significant, it allows
an unbounded upper range for $\gamma$. 

A natural choice for $m=2, N=4$ is to consider the potential for a
coupled quartic nonlinear oscillators:
\be
V= \sum_{i=1}^2 \left [ -2\omega_i^2 z_i^2 - \frac{\alpha_i}{4} z_i^4 \right ]
- \beta z_1^2 z_2^2.
\label{potv}
\ee
\noindent There are four distinct regions in parameter-space for which exact
solutions exist\cite{ml}. For example, for the choice of the parameters
$\omega_1=\omega_2\equiv \omega_0, \alpha_1=\alpha_2\equiv \alpha, 2\beta =
3 \alpha$, $z_i$ satisfy the equation:
\be
\ddot{z}_1 + \omega^2 {z}_1 + \alpha ( z_1^3 + 3 z_1 z_2^2) = 0,\
\ddot{z}_2 + \omega^2 {z}_2 + \alpha ( z_2^3 + 3 z_1^2 z_2) = 0, \
\omega^2=4(\omega_0^2 -\gamma^2).
\ee
\noindent These two equations are separable in the co-ordinate,
$u=z_1+z_2, v=z_1-z_2$ and describe two cubic oscillators:
\be
\ddot{u} + \omega^2 u + \alpha u^3=0, \
\ddot{v} + \omega^2 v + \alpha v^3=0.
\label{cubic1}
\ee
\noindent As discussed for the case of $m=1,N=2$, there are three distinct
regions in the parameter-space for which exact solutions exist for a cubic
oscillators. In each region,the solutions for $z_1(t)$ and $z_2(t)$ may be
constructed by combining $u\equiv u(A_1,k_1)$ and $v \equiv v(A_2, k_2)$ for
different values of the amplitude $A$ and the modulus $k$. In particular,
\be
z_1(t)=\frac{1}{2} \left [ u(A_1,k_1) + v(A_2,k_2) \right ],
z_2(t)=\frac{1}{2} \left [ u(A_1,k_1) - v(A_2,k_2) \right ].
\ee
For example, we may consider the solution in the range $\omega>0$, $\alpha>0$.
In this range the solutions for $u$ and $v$ are given as:
\bea
u_i= A_i \ cn (\Omega_i t,k_i),\ \ \ \Omega_i=\sqrt{\omega^2 + \alpha A_i^2}, \ \
k_1^2=\frac{\alpha A_1^2}{2 \Omega_1^2},\ \ \ \ i=1,2,\ \ u_1=u,u_2=v.
%v= A_2 \ cn (\Omega_1 t,k_2),\ \ \ \Omega_2=\sqrt{\omega^2 + \alpha A_2^2}, \ \ \
%k_2^2=\frac{\alpha A_2^2}{2 \Omega_2^2}.
\eea
Therefore , we have:
\bea
z_1(t)=A_1 \ cn (\Omega_1 t,k_1)+ A_2 \ cn (\Omega_1 t,k_2)\\
z_2(t)=A_1 \ cn (\Omega_1 t,k_1)- A_2 \ cn (\Omega_1 t,k_2)
\eea
and
\bea
z_1^+=\sum_{i=1}^{2} E(A_i,k_i), \ \ z_2^+=\sum_{i=1}^{2} (-1)^{i+1}E(A_i,k_i)
\eea
where
\bea
E(A_i,k_i)=2 \gamma \left[ \frac{\cos^{-1} \{ dn(\Omega_i t,k_i) \}
sn(\Omega_i t,k_i)}{\Omega_i\sqrt{1-dn^2(\Omega_i t,k_i)}}\right].
\eea
%\bea
%z_1^+= 2 \gamma \left[ \frac{\cos^{-1} \{ dn(\Omega_1 t,k_1) \}
%sn(\Omega_1 t,k_1)}{\Omega_1\sqrt{1-dn^2(\Omega_1 t,k_1)}}+\frac{\cos^{-1} \{ dn(\Omega_2 t,k_2) \}
%sn(\Omega_2 t,k_2)}{\Omega_2 \sqrt{1-dn^2(\Omega_2 t,k_2)}}\right]\nonumber\\
%z_2^+= 2 \gamma \left[ \frac{\cos^{-1} \{ dn(\Omega_1 t,k_1) \}
%sn(\Omega_1 t,k_1)}{\Omega_1\sqrt{1-dn^2(\Omega_1 t,k_1)}}-\frac{\cos^{-1} \{ dn(\Omega_2 t,k_2) \}
%sn(\Omega_2 t,k_2)}{\Omega_2 \sqrt{1-dn^2(\Omega_2 t,k_2)}}\right].
%\eea
In terms of $x_i, i=1,2,3,4$ variables the solutions are:
\bea
x_1&=&\sum^{2}_{i=1}\frac{1}{\sqrt{2}}\left[E(A_i,k_i)+A_i \ cn (\Omega_i t,k_i)\right],\ \ \ 
x_2=\sum^{2}_{i=1}\frac{1}{\sqrt{2}}\left[E(A_i,k_i)-A_i \ cn (\Omega_i t,k_i)\right],\nonumber\\
x_3&=&\sum^{2}_{i=1}(-1)^{i+1}\frac{1}{\sqrt{2}}\left[E(A_i,k_i)+A_i \ cn (\Omega_i t,k_i)\right],\nonumber\\
x_4&=&\sum^{2}_{i=1}(-1)^{i+1}\frac{1}{\sqrt{2}}\left[E(A_i,k_i)-A_i \ cn (\Omega_i t,k_i)\right],
\eea
 Non-singular stable solutions are obtained for $0< k_1 < 1$ and $0< k_2<1$. Solutions for the other range of the
parameters are obtained similarly by using the Eqs. (\ref{solex2}) and (\ref{solex3}). For the choice of the parameters
$\omega_i=\omega$, $\alpha_i=\alpha$ and $\beta=0$, the potential in Eq. (\ref{potv}) becomes spherically
symmetric. This case will be considered in the next section where spherically symmetric systems are discussed.

\subsubsection{Embedded Rotational Symmetric System}
 
The original problem involving $2m$ particles  has been reduced to the
study of a $m$ particle sub-system in terms of coordinates $z_i$. Rotationally
symmetric potential in this $m$ dimensional subspace is considered. 
A new variable is introduced,
\be
r^2 = \sum_{i=1}^m z_i^2, \ 
\ee
\noindent which may be identified as a radial variable in $m$ dimensional
Euclidean sub-space spanned by $z_i$ co-ordinates. The $z_i$ co-ordinates
satisfy the equation,
\be
\ddot{z}_i - 4 \gamma^2 z_i -  \frac{1}{r} \frac{\partial V}{\partial r}
z_i =0
\label{spheri-eqn}
\ee
\noindent for rotationally symmetric potential $V(r)$. This equation can be
expressed solely in terms of radial variable in $m$ dimensional
hyper-spherical co-ordinate,
\be
\ddot{r} - 4 \gamma^2 r - \frac{L^2}{r^3} -
 \frac{\partial V}{\partial r}=0,
\ee
\noindent where $L^2$ is the square of the angular momentum and a constant
of motion. Exact solutions may be obtained for specific choices of $V(r)$.
The potential is chosen as,
\be
V(r)=-2\omega_0^2 r^2 - \frac{\alpha}{4} r^4 
-\frac{\delta}{2 r^2}, \ \alpha, \delta \in \Re,
\label{V}
\ee
\noindent for which Eq. (\ref{spheri-eqn}) reduces to the form: 
\be
\ddot{r} + \omega^2 r + 
\left ( \alpha r^3 - \frac{L^2 - \delta}{r^3} \right ) = 0, \ \
\omega^2 \equiv 4 (\omega_0^2 - \gamma^2 ).
\label{final} 
\ee
\noindent 
%In contrast to the coupled
%linear chain of oscillators with uniform gain-loss parameters studied in
%Ref. \cite{ben1}, the region in the parameter space for which stable classical
%solutions exist remains unchanged for the the system under investigation,
%even  if the total number of oscillators is increased. 

Three different parametric choices
of the potential are discussed below separately:\\

(a) $\alpha=\delta=0$: This describes a trivial generalization of coupled
dissipative oscillator model considered in Ref. \cite{ben}. In fact, the
system governed by the Hamiltonian (\ref{ham-z}) for this case is m copies
of the system considered in Ref. \cite{ben} with $\epsilon=-\omega_0^2$. Each
of the $m$ uncoupled sub-systems describes a Hamiltonian of two oscillators
with balanced loss and gain. A shift of the co-ordinates $z_i^-$ by an amount
$\frac{2 \gamma \Pi_i}{4 \omega^2}$ casts the governing equations of motion
as $m$ decoupled harmonic oscillators, which are exactly solvable.

(b) $\alpha‌\ne0,\delta \ne 0$: 
In this case the Eq. (\ref{final}) becomes:

\be
\ddot{r} + \omega^2 r + 
\left ( \alpha r^3  - \frac{L^2-\delta}{r^3} \right ) = 0,
\label{final1} 
\ee
The solution of which may be written as:

\bea
r(t)&=&\left[\alpha_3-\left(\alpha_3-\alpha_2\right)sn^2(\lambda t,k)\right]^{\frac{1}{2}},
\ \ 0<k^2=\frac{\alpha_3-\alpha_2}{\alpha_3-\alpha_1}<1,\nonumber\\
&&\ \ \ \ \lambda^2=\frac{\alpha}{2}(\alpha_3-\alpha_1),
\label{rr}
\eea
where $\alpha_1, \alpha_2,\alpha_3$ are constants satisfying the conditions:
\bea
&&\alpha_1+\alpha_2+\alpha_3=-\frac{2 \omega^2}{\alpha},\ \ \alpha_1\alpha_2\alpha_3=-\frac{2(L^2-\delta)}{\alpha},\\
&&\alpha_1\alpha_2+\alpha_2\alpha_3+\alpha_3\alpha_1=\frac{4 E}{\alpha}.
\eea
Further, Eq. (\ref{rr}) implies that for real $r$, $\alpha_2$ must be positive and the restriction on
$k$ implies $\alpha_2>\alpha_3, \alpha_1$ and $E$ is the energy corresponding to $m$ oscillators. The angular momentum $L$ is a 
constant of motion. In general the solution for the angular part is a bit involved and is discussed in 
Appendix-B. However, a simplified solution for the angular variables may be obtained if we take the angular
momentum to be identically zero for which all the angular variables become constant.
With this simplifying assumption, all the $z_i$ are easily determined and we have:
\bea
z_i&=&\left[\alpha_3-\left(\alpha_3-\alpha_2\right)sn^2(\lambda t,k)\right]^{\frac{1}{2}} f_i(\theta_i),
\eea
where $f_i(\theta_i)$ is the function of angular coordinates of the form appears
in the hyper-spherical coordinate system and is a constant in this case.
The solutions of $z_i^+$ are of the form
\be
z_i^{+}=2\gamma f_i(\theta_i) I(t),
\ee 
with
\bea
I(t)=-\frac{i\left(\alpha_3 w^2 EllipF(l(t),m)+(\alpha_3-\alpha_2)EllipPi(\frac{1}{w^2},l(t),m)
\right)n(t)}{w \lambda dc(\lambda t,k)[\alpha_3-(\alpha_3-\alpha_2)sn^2(\lambda t,k)]^{\frac{1}{2}}},
\eea

\bea
l(t)&=&i \sinh^{-1}{(w sc[\lambda t,k])},\ \ w^2=\frac{\alpha_2}{\alpha_3},\ \ 0<m=\frac{w^2}{(1-k)}<1,\\
n(t)&=&[(1+w^2 sc^2[\lambda t,k])(1-k sn^2[\lambda t,k])]^{\frac{1}{2}}.
\eea
It should be mentioned here that the expression of $I(t)$ encounters singularity and therefore the 
solutions of $z_i^+$'s are not stable.

%region of stable solution in the parameter-space
%remains same for any number of particles which is markedly different to the example
%considered in Ref. \cite{ben1} where the stability region disappears as the number of oscillators
%is increased.

(c) $\delta=0$: The system reduces to a chain of coupled nonlinear
oscillators with non-linearity arising due to cubic terms. This may be considered
as the rotationally symmetric case of Eq. (\ref{cubic1}) in $m$ dimension. 
The Hamiltonian which produces
Eq. (\ref{final}) is known to be integrable for several distinct choices of
the parameters\cite{ml}. 
For $\delta=0$ and m=2, Eq. (\ref{final1}) describes a quartic oscillator and the solution for
$r$ takes the following form:
\bea
r(t)=[\frac{2}{\alpha}(E+\omega^2-\sqrt{\frac{\alpha}{2}\omega})]^{\frac{1}{2}}\sin{(\frac{\alpha}{2}(t+c))}
\eea
where $c$ is a constant of integration and $z_1$, $z_2$ are respectively given as:
\bea
z_1&=&[\frac{2}{\alpha}(E+\omega^2-\sqrt{\frac{\alpha}{2}\omega})]^{\frac{1}{2}}\sin{(\frac{\alpha}{2}(t+c))}\cos{\theta_1},\\
z_2&=&[\frac{2}{\alpha}(E+\omega^2-\sqrt{\frac{\alpha}{2}\omega})]^{\frac{1}{2}}\sin{(\frac{\alpha}{2}(t+c))}\sin{\theta_1}
\eea
and the form of $z_1^+$ and $z_2^+$ are obtained as:

\bea
z_1^+&=&-2\gamma(\frac{\alpha}{2}[\frac{2}{\alpha}(E+\omega^2-\sqrt{\frac{\alpha}{2}\omega})]^{\frac{1}{2}}\cos{(\frac{\alpha}{2}(t+c))}\cos{\theta_1},\\
z_2^+&=&-2\gamma(\frac{\alpha}{2}[\frac{2}{\alpha}(E+\omega^2-\sqrt{\frac{\alpha}{2}\omega})]^{\frac{1}{2}}\cos{(\frac{\alpha}{2}(t+c))}\sin{\theta_1}.
\eea
It should be mentioned here that the region of stable solution in the parameter-space
remains same for any number of particles which is markedly different to the example
considered in Ref. \cite{ben1} where the stability region disappears as the number of oscillators
is increased.
The solutions for $r$ and angular variables $\theta_i, i=1,2,....,m-1$ for nonzero $L$ are discussed in the Appendix B.

%Constants of motion for $\Gamma_i=1, \forall i$:
%\be
%\Pi_i= \dot{z}_i^+ + 2 \gamma z_i^-= 2 P_{z_i^+} + \gamma z_i^{-},\ \
%I_i= \sum_{\substack{j=1\\ j \neq i}}^m \left ( z_i^- \dot{z}_j^- - z_j^- 
%\dot{z}_i^-\right )^2, \ \ i=1, 2, \dots m
%\ee 
 
%(c) $\alpha=0$ models of quasi-exactly solvable model.

\subsubsection{Calogero-type systems with four-body interaction}

In this subsection a solvable many particle model with four-body interaction
in one dimension in the presence of balanced loss and gain terms is investigated. 
Many-body systems with four-body interaction
have been considered earlier in the literature \cite{wolf,hack,bac}. 
The exactly solvable quantum models of Calogero and Sutherland type
with translationally invariant two and four-body interactions is investigated in Ref. \cite{hack}. An 
exactly solvable four-body interaction with non-translationally invariant interactions is discussed in
Ref. \cite{bac}. In our example, the four-body inverse square interaction is generated in the 
 presence of balanced loss and gain terms in the original coordinate $x_i$ by considering a Calogero-type 
of potential for the reduced system in $z_i$ coordinates. The exact known solutions of Calogero-type system
is then exploited to obtain the exact solutions of the four-body model.

The Calogero-type of system arises if we take the potential $V(z_i)$ as
\be
V(z_i)=-\sum_{i}^m 2\omega_0^2 z^2_i-\sum_{\substack{i,j=1 \\ i < j}}^m\frac{g^2}{2(z_i-z_j)^2},
\ee
and the Eq. of motion (\ref{homo-eqn}) becomes:
\bea
&& \ddot{z}_i +\omega^2 {z}_i -\sum_{j,(j\ne i)}^m\frac{g^2}{(z_i-z_j)^3}= 0 
\label{h},\\
&& z_i^+(t) = 2 \gamma \int z_i(t) dt + C_i, \ \ i=1, 2, \dots m.
\label{homo-eqn1}
\eea
The four-body interaction manifests itself in the potential as well as in the 
equations of motion, when $z_i$'s  are replaced by the original coordinates $x_i$.
In particular, the potential 
\bea
V=-\sum_{i}^m \omega_0^2 (x_{2i-1}-x_{2i})^2-\sum_{\substack{i,j=1 \\ i < j}}^m\frac{g^2}{(x_{2i-1}-x_{2i}-x_{2j-1}+x_{2j})^2},
\label{vfour}
\eea
describes a pair-wise two-body harmonic term plus a four-body inverse square
interaction. 
The equations of motion are,
\bea
&&\ddot{x}_{2l-1}-2\gamma \dot{x}_{2l-1}+2\omega_0^2(x_{2l-1}-x_{2l})-\sum_{\substack{i=1\\i\ne l}}^m\frac{2g^2}{(x_{2i-1}-x_{2i}-x_{2l-1}+x_{2l})^3}=0,\nonumber\\
&& \ddot{x}_{2l}+2\gamma \dot{x}_{2l}-2\omega_0^2(x_{2l-1}-x_{2l})+\sum_{\substack{i=1\\i\ne l}}^m\frac{2g^2}{(x_{2i-1}-x_{2i}-x_{2l-1}+x_{2l})^3}=0,\  l=1,..m.
\label{nm1}
\eea
 Note that the permutation symmetry of the reduced system in terms of $z_i$ is
lost when expressed in terms of the original coordinates $x_i$.

The classical solutions of Eq. (\ref{h}) are the well known solutions
of rational Calogero system with a harmonic confinement and the solutions 
for $ z_i^+(t)$ can be obtained by integrating Eq. (\ref{homo-eqn1}).
The solution of Eq. (\ref{h}) may be obtained
from the Lax-pair formulation\cite{sata} (see Appendix-C), and is given by
the eigenvalue of the following matrix: 
\bea
Q(t) = Q(0) \cos{(\omega t)} + \omega^{ -1} \dot{Q}(0) \sin{(\omega t)},
\label{q}
\eea
where $Q(0)=X(0)$ and $ \dot{Q}(0)=L(0)$ are obtained from the initial values of $z_i(0)$ and $p_{z_ i}(0)$
(for the expression of the matrices Q, L, ${\cal M}$, see Appendix-C).
The expressions for $z_i^+$'s  are given by the eigenvalues of the matrix $R(t)$, where 
\bea
R(t)=2\gamma \left(\frac{X(0)}{\omega} \sin{(\omega t)} - \omega^{ -2} L(0) \cos{(\omega t)}\right).
\label{r}
\eea
The closed form expressions for the eigenvalues of $Q(t)$ and $R(t)$ can be obtained easily for small $N$. 

The case $m=1,N=2$ describes a system with only two-body interaction and is discussed in detail in Ref. \cite{ds-pkg}.
The four-body interaction appears in the Hamiltonian for $N\ge4$. 
For the case $N=4$, the eigenvalues of $Q(t)$ gives:
\bea
z_1&=&\frac{1}{2}\left[(a_1+a_2)+\sqrt{(a_1-a_2)^2+b^2}\right]\\
z_2&=&\frac{1}{2}\left[(a_1+a_2)-\sqrt{(a_1-a_2)^2+b^2}\right]
\eea
where
\bea
a_j&=&z_j(0)\cos{\omega t}+\frac{p_{z_j}(0)\sin{\omega t}}{\omega},\ \ j=1,2\\
b^2&=&\frac{g^2\sin^2{\omega t} }{(z_1(0)-z_2(0))^2}.
\eea
and from Eq. (\ref{r}), the eigenvalues of $R(t)$ may be written as
\bea
z_1^+&=&\frac{\gamma}{\omega}\left[(c_1+c_2)+\sqrt{(c_1-c_2)^2+d^2}\right]\\
z_2^+&=&\frac{\gamma}{\omega}\left[(c_1+c_2)-\sqrt{(c_1-c_2)^2+d^2}\right]
\eea
where
\bea
c_j&=&z_j(0)\sin{\omega t}-\frac{p_{z_j}(0)\cos{\omega t}}{\omega},\ \ j=1,2\\
d^2&=&\frac{g^2\cos^2{\omega t} }{(z_1(0)-z_2(0))^2}.
\eea
Now $x_i,i=1,2,3,4$ can easily be determined from Eq. (\ref{trans}). The solutions are periodic and stable.
Solutions for $N>4$ may be obtained in a similar way.\\

A few comments are in order before concluding the section;\\

i) The most general four-body inverse-square interaction for a many-particle system
 is of the form\cite{wolf}:
\be
V_4=\sum_{\substack{i,j,p,q=1 \\ i \ne j\ne p\ne q }}^N\frac{g}{(x_i-x_j-x_p+x_q)^2}
\label{V4}
\ee
which is invariant under permutation symmetry. On the other hand, the potential in Eq. (\ref{vfour}) 
lacks permutation symmetry. It is interesting to note that the inverse-square part of the 
potential in Eq. (\ref{vfour}), i.e.
\bea
\tilde{V}_4=\sum_{\substack{i,j=1 \\ i < j}}^m\frac{g^2}{(x_{2i-1}-x_{2i}-x_{2j-1}+x_{2j})^2},
\label{vfour1}
\eea
forms a subset of the set of all the terms in (\ref{V4}). Further, although the form of $\tilde{V}_4$ changes with a 
change in the  representations of $M,D$ and $R$, it always generates terms which form a subset of terms 
presented in $V_4$.

ii) It should be mentioned here that another four-body exactly solvable model may be generated
in a similar way if we consider the Sutherland model for the reduced system in $z_i$ coordinate.
However, in this case in order to remove the harmonic term proportional to $4\gamma^2$ a counter term
should be included in the potential, i. e., the potential should be of the form:
\be
V(z_i)=-\sum_{i}^m 2\gamma^2 z^2_i-\sum_{\substack{i,j=1 \\ i < j}}^m\frac{g^2}{\sin^2{g(z_i-z_j)}}.
\ee
This kind of potential gives Sutherland model for the reduced system in $z_i$ coordinates whose 
classical solutions are well known\cite{sata}. These solutions may be exploited to generate the 
solutions of a system having trigonometric four-body interaction. The solution of
 Calogero model for any root system with four-body interaction can be obtained in
a similar manner. For an example $D_N$ type Calogero model with four-body interaction 
is carried out in Appendix-C.

\subsection{Rotationally invariant potential}
 
A constant of motion may be constructed for a class of potential
$V$: 
\bea
\hat{L} & = & \frac{1}{2} \left ( P^T D X + X^T D P \right )\nonumber \\
& = & \tilde{X}^T Q \tilde{P},
\eea
\noindent which can be interpreted as angular momentum in a co-ordinate system
endowed with the metric $M_d$ for specific forms of $R$.
\noindent The time-variation of $\hat{L} $ is related to the potential $V$,
\be
\frac{d\hat{L} }{dt}= - X^T D \frac{\partial V}{\partial X},
\ee
\noindent and $\hat{L} $ is a constant of motion provided the following condition is
 satisfied:
\be
X^T D \frac{\partial V}{\partial X} =0.
\label{con-int}
\ee
\noindent The constraint (\ref{con-int}) may be solved with the following 
ansatz for $V$:
\be
V\equiv V(r), \ \ r^2 \equiv X^T G X, \ \ \{G, D\}=0,
\ee
\noindent where $G$ is a symmetric matrix. Both $M$ and its inverse $M^{-1}$
are symmetric and anti-commute with $D$. For $N=2$, with the representation
of the matrices $M, R, D$ given by Eq. (\ref{choice0}), $G$ is uniquely fixed
to be $G=M=\sigma_x$. For $N > 2$, one may choose $G= a M + b M^{-1}, \ a, b
\in \Re$. However, this choice is not unique, there are several other
possibilities. For example, the matrices $M, R. D$ may be chosen from a
real representation of Clifford algebra\cite{clif} for $N=2^m, m \in \mathbb{Z^+}$. It is
always possible to find a representation where the number of non-diagonal
symmetric matrices is $m$. Thus, for $N >2$, $G$ and $M$ can be chosen
independently. 

\subsubsection{Exactly Solvable Models with $N=2$}

The existence of two first integrals $H$ and $\hat{L} $ implies that
the governing system is integrable for $N=2$ for any $V(r)$. The 
representations of $M, R, D$ are given by Eq. (\ref{choice0}) and the matrix
$G=M=\sigma_x$ and $r=\sqrt{x_1 x_2}$.  There are several choices of $V$ for
which exactly solvable models can be constructed. The example of a system of
coupled nonlinear oscillators is considered in this article in some detail. 
The potential is chosen as,
\be
V(r) = \frac{1}{2} \omega^2 r^2 + \frac{\alpha}{4} r^4, \ \
\label{pot-no}
\ee
\noindent which gives the following Eq. of motion,
\be
\ddot{x}_i - (-1)^{i+1} 2 \gamma \dot{x}_i +\omega^2 x_i + 
\alpha \left ( x_1 x_2 \right ) x_i=0, \ i=1, 2.
\label{noscN2}
\ee
\noindent The parameter $\alpha$ controls the strength of the nonlinear
interaction term, whereas $\omega$ is the angular frequency of the harmonic
term.  A system of uncoupled oscillators with balanced loss and gain, i.e.,
the system considered in Ref. \cite{bat}, is obtained for $\alpha=0$.
\noindent The constant of motion $\hat{L} $ has the form,
\be
\hat{L} = (\frac{-\gamma}{2}) \left ( \dot{x}_1 x_2 - \dot{x}_2 x_1 -
2 \gamma x_1 x_2 \right ),
\ee
\noindent which is independent of $\alpha$. The overall multiplication factor
has no significance and one may define $\tilde{L}=-\frac{2}{\gamma} \hat{L} $ as the
constant of motion. The expression for $\tilde{L}$ may also be obtained directly
from Eq. (\ref{noscN2}). The constant value of $\hat{L} $ is
determined from the initial conditions imposed on $x_i(0)$ and $\dot{x}_i(0)$.
The value $\hat{L} =0$ may be obtained for different sets of boundary conditions. For
$\hat{L} =0$, $x_1(t)$ and $x_2(t)$ are related to each other. In particular,
\be
x_1= c_0 e^{\gamma t} q(t), \ \
x_2= d_0 e^{-\gamma t} q(t), \ \ c_0, d_0 \in \Re,
\ee
\noindent where $q(t)$ satisfies the equation of a cubic nonlinear oscillator:
\be
\ddot{q} + \Omega^2 q + \Gamma q^3, \ \
\Omega^2 \equiv \left ( \omega^2 - \gamma^2 \right ), \ \ 
\Gamma \equiv \alpha c_0 d_0.
\ee
\noindent which is exactly solvable. This equation is identical to the second
Eq. of (\ref{cubic-o}) with the identification $\Omega^2=\omega^2,
\Gamma=\alpha$. Exact non-singular solutions for $q$ can be found. However,
$x_1$ is always a decaying solution, while $x_2$ is a growing solution.
The system of harmonic oscillators with balanced loss and gain and without
any coupling between the two has similar solutions. The introduction of a
nonlinear coupling between the two as specified in Eq. (\ref{noscN2}) do
not give any stable solutions. The specific type of non-linearity for which
$\hat{L} $ is a constant of motion is not suitable for obtaining classically stable
solutions.

\subsubsection{Solvable Model of $N >2$ }

A solvable model of $N=2 m$ numbers of coupled nonlinear oscillators are
presented in this section. 
The matrices $M$, $D$, $R$ are given by Eq. (\ref{choice1}) with  $\eta= 
\gamma I_m$ and  $\kappa=I_m$ and $G$ is chosen
as $G=M^{-1}$ so that the variable $r$ has the following form:
\be
r^2= \sum_{i=1}^{m} x_{2 i-1} x_{2 i} = \sum_{i=1}^m \left [ \left ( z_i^+
\right )^2 - \left ( z_i^- \right )^2 \right ].
\label{N-radi}
\ee
\noindent It may be noted that $r$ has the interpretation of the radial
variable in a pseudo-Euclidean co-ordinate with the signature of the metric
as $(1,-1,1,-1, \dots, 1, -1)$. With the
choice of the potential $V \equiv V(r)$, the Hamiltonian (\ref{ham-z}) with
$\gamma=0$ is rotationally invariant in this pseudo-Euclidean co-ordinate.
The rotational invariance of the Hamiltonian (\ref{ham-z}) is partially lost
for $\gamma \neq 0$, since the term linear in $\gamma$ is the sum of angular
momenta for rotations in $m$ planes specified by `$z_i^--z_i^+$'. The 
Hamiltonian (\ref{ham-z}) is invariant under rotation when the planes of
rotations are chosen as `$z_i^--z_i^+$'

The Eqs. of motion for a $V \equiv V(r)$  with $r$
given by Eq. (\ref{N-radi}) reads,
\be
\ddot{z}_i^+ - 2 \gamma \dot{z}_i^- + 
\frac{1}{r} \frac{\partial V}{\partial r} z_i^+=0, \
\ddot{z}_i^- - 2 \gamma \dot{z}_i^+ +
\frac{1}{r} \frac{\partial V}{\partial r} z_i^-=0
\ i=1, 2, \dots m.
\label{pot-rot-eqn}
\ee
\noindent Multiplying the Eq. for $z_i^+$ by $\dot{z}_i^-$,  Eq. for $z_i^-$
by $\dot{z}_i^+$ and subtracting the resulting equations, $m$ constants of
motion may be obtained as,
\be
\hat{L} _i = \dot{z}_i^+ z_i^- - z_i^+ \dot{z}_i^- + \gamma \left [ ( z_i^+)^2
-(z_i^-)^2 \right], \ i=1, 2, \dots, m.  
\ee
\noindent These constants of motion along with the Hamiltonian constitute
$m+1$ number of integrals of motion,
\be
\{\hat{L} _i, \hat{L} _j \}_{PB} =0, \ \{H, \hat{L} _i\}_{PB}=0,
\ee
\noindent implying that the system at least partially integrable.
The co-ordinates may be parametrized in terms of $m$ functions
$q_i(t)$ as,
\be
z_i^+(t) = q_i(t) \cosh(\gamma t), \
z_i^-(t) = q_i(t) \sinh(\gamma t),
\ee
\noindent for which the values of all the constants of motion are zero, i,e,
$\hat{L}_i=0, \ \forall \ i$ and $r^2=\sum_{i=1}^m q_i^2\equiv q^2$. With the
substitution of $z_i^{\pm}(t)$ in the equations of motion (\ref{pot-rot-eqn}),
the following equations in terms of $q_i$'s are obtained:
\be
\ddot{q}_i - \gamma^2 q_i + 
\frac{1}{q} \frac{\partial V(q)}{\partial q} q_i=0,
i=1, 2, \dots, m.
\label{qq-eqn}
\ee
\noindent
\noindent Exactly solvable models may be constructed for suitable choices
of $q$. 

A particular example of exactly solvable model is a chain of coupled nonlinear
oscillators. The equation (\ref{qq-eqn}), with the choice of the potential as
in (\ref{pot-no}) and the radial variable defined by (\ref{N-radi}), can be
expressed as,
\be
\ddot{q}_i +  \Omega^2 q_i + \alpha q^2 q_i=0, \ \Omega^2=\omega^2 - \gamma^2.
\label{hyperbolic}
\ee
\noindent which is exactly solvable. In terms of $r$ in Eq. (\ref{N-radi}),
 Eq. (\ref{hyperbolic}) takes the form of Eq. (\ref{final}) with $\delta=0$ and
$\omega$ replaced by $\Omega$. Therefore, with this identification of 
parameters the solutions of Eq. (\ref{final}) are also valid solutions of 
 Eq. (\ref{hyperbolic}).
 The solutions of $x_{2i}$ are always a
decaying one, while that of $x_{2i-1}$ are a growing one.
The system of harmonic oscillators with balanced loss and gain and without
any coupling between them has similar solutions. The introduction of a
nonlinear coupling as in (\ref{pot-no}) does not produce any stable solutions.

\section{Summary \& Discussions}

A Hamiltonian formulation of a generic many-body system with balanced
loss and gain has been investigated. The form of a generic many-particle
Hamiltonian including gauge potential has been assumed. The requirement that the resulting
Eqs. of motion contain balanced loss and gain terms severely restricts the 
allowed range of the gain-loss coefficient and the possible form of the 
interacting potential. The balancing of the loss and gain terms occur in a 
pair-wise fashion, i.e corresponding to the coefficient $\gamma$ of a loss
term there exits necessarily a gain term having the coefficient $-\gamma$ and
balancing of $\gamma$ by two of more terms are not allowed. It has been found that
some of the well-known many-particle systems like Calogero models are not 
amenable to Hamiltonian formulation for three or more particles, if they are
generalized to include balanced loss and gain terms. In the Calogero model
each particle interacts with rest of the particles which is manifested in the 
respective equations of motion. If this feature is sacrificed, then a Hamiltonian
formulation for many-body system having inverse square potential plus a harmonic
term may be incorporated in the context where loss and gain are balanced. In fact,
in spite of the imposing restriction, the Hamiltonian formulation for a large class of
system is formulated with balancing loss and gain terms that also includes some of the 
known examples. Finally, it is shown that the Hamiltonian can always be reformulated
 in the background of a pseudo-Euclidean metric through some transformations of the
 coordinate.  A Hamiltonian formulation for systems with space-dependent balanced 
loss and gain terms is also presented.
The system having balanced loss and gain is investigated from the viewpoint of exactly
solvable models. Two types of exactly solvable models with balanced loss and gain 
 are considered. Type-I system is characterized by a potential which has a translational 
symmetry although the Hamiltonian does not respect this symmetry. The type-II system
is characterized by a potential which has a rotational symmetry in the background of a 
space having pseudo-Euclidean metric of the form $g_{ij}=(-1)^{i+1}\delta_{ij}$.  
The Hamiltonian for type-II models is not invariant under rotation in any plane,
but, invariant under rotations in specific planes. For both 
types of systems, apart from the Hamiltonian, $m$ number of integrals of motion
are constructed where $N=2m$ is the number of particles in the system.These integrals of
motion are in involution implying that the system is at least partially
integrable. The existence of $m$ degrees of freedom make it possible to determine the dynamics
of the original system with $2m$ degrees of freedom in terms
of an effective system having $m$ degrees of freedom. The
exactly solvability of the effective system ensures the same for the original
Hamiltonian. 

For the type-I models, stable classical solutions are obtained in
terms of Jacobi elliptic functions for particular ranges of parameters. 
For the case of a single quartic nonlinear oscillator, stable
solutions are obtained even if the gain-loss parameter $\gamma$ is varied
without any upper bound. Exact solutions are obtained in closed 
analytical form for several models which are appropriate generalizations 
of well-known systems like, two particles coupled nonlinear oscillators, 
Henon-Heils system, Calogero-type model etc. 
Some exactly solvable many particle systems with balanced loss and gain 
and are interacting via a four-body potential are investigated.
The exact solutions of these model are obtained by exploiting the known solutions
 of Calogero-type of models. It is found that the balancing of the loss and gain terms in
a pair-wise fashion highly restricts the possible form of the four-body interaction.
Exact solutions are obtained for several type-II models, including coupled
chain of nonlinear oscillators. However, the solutions are not stable and there
exists no region in the parameter-space for which stable solutions are possible.
There are equal number of growing and decaying solutions for even
number of particles. Exact solutions are obtained in terms of Jacobi elliptic
functions multiplied by an exponentially decaying/growing factor. 

The Schwinger-Keldysh formalism is an useful tool for studying
nonequlibrium many-particle systems.  It is being used in a variety
of contemporary topics like driven open quantum systems\cite{diehl,diehl1,jsm},
time-dependent density-functional theory\cite{leewuen}, relativistic
hydrodynamics, physics of black-holes, dynamics of entanglement in quantum
field theory etc.\cite{mr}. The invariance of Schwinger-Keldysh action under
time-reversal symmetry plus time-translation corresponds to thermodynamic
equilibrium\cite{diehl1}. The Hamiltonian is the generator of the
time-translation and unitary time evolution, while invariance under
time-reversal symmetry conforms to principle of detailed balance.

The above result is worth comparing with that of the systems with balanced
loss and gain, where an equilibrium is reached in regard to energy transfer
between the system and the bath for unbroken ${\cal{PT}}$ 
symmetry\cite{ben,ben1,ds-pkg}. The action corresponding to the
Lagrangian in Eq. (\ref{lag}) and the associated Hamiltonian are invariant
under time-translation, as in the case of Schwinger-Keldysh action at
thermodynamic equilibrium. An apparent difference between the two cases arises
in respect to discrete symmetries. The Schwinger-Keldysh action at
thermodynamic equilibrium is invariant under time-reversal symmetry, whereas
unbroken ${\cal{PT}}$ symmetry is essential for the existence of stable
classical solutions of $H$ in Eq. (\ref{Ham}) or the respective quantum bound
states\cite{ben,ben1,ds-pkg, dsp5}. This apparent difference may be removed,
if a non-conventional time-reversal symmetry\cite{nctrs} is used instead of
the conventional one, which generates the same canonical transformations as
in the case of ${\cal{PT}}$ symmetry. For example, the Hamiltonian in Eq.
(\ref{ham-z}) can be interpreted as a system of $m$ particles on the
two dimensional plane embedded in a three dimensional system. Following Ref.
\cite{nctrs}, a non-conventional time-reversal symmetry($\hat{\cal{T}}$) for
the many-particle system may be defined as,
\bea
&& \hat{\cal{T}}= exp(i \pi \sum_{i=1}^m J_{z_i^-}) \ {\cal{T}},\nonumber \\
&& \hat{\cal{T}}: z_i^+ \rightarrow - z_i^+, \ z_i^- \rightarrow z_i^-, \
P_{z_i^+} \rightarrow P_{z_i^+}, \ \ P_{z_i^-} \rightarrow - P_{z_i^-},
\eea
\noindent where $J_{z_i^-}$ denotes generator of rotation for the $i$-th
particle around $z^-$-axis and ${\cal{T}}$ is the conventional time-reversal
operator. If the third degree of freedom for the $i$-th
particle is denoted as $z_i^0$, then $\hat{\cal{T}}: z_i^0 \rightarrow -
z_i^0, \ P_{z_i^0} \rightarrow P_{z_i^0}$ ensures that $\hat{\cal{T}}
{\cal{T}}^{-1}$ describes a proper rotation in three dimensional space.
Further, for any two quantum mechanical states $|\psi\rangle$ and
$|\phi\rangle$, $\langle \hat{\cal{T}} \phi|\hat{\cal{T}} \psi \rangle=
\langle \psi|\phi\rangle$ signifies time-reversal invariance. Thus, the
${\cal{PT}}$ symmetry on the plane and the non-conventional time-reversal
symmetry $\hat{\cal{T}}$ in three dimensions induces the same canonical
transformations on the $z^--z^+$ plane. Consequently, the conditions for
thermodynamic equilibrium of the Schwinger-Keldysh action and equilibrium
condition for systems with balanced loss and gain may be identified as
similar.\\

\noindent Some future directions of study may be listed as follows:
\begin{itemize}

\item The many particle interaction of Calogero model is governed by
root system of finite reflection groups. The type of Calogero model considered in this
work belongs to A-type root system. However, Calogero-Moser system is integrable for
other root systems such as  B,C,D and also for the exceptional 
and non-crystallographic root systems. Therefore, the investigation of integrability and exact solvability of all
such root systems in the presence of the balanced loss and gain terms will be a part of the future studies. Further,
the integrability and exact solvability of impenetrable system where the nearest neighbour and the next-to-nearest 
neighbours interact\cite{jain-khare}  in the presence of the balanced loss and gain terms will also be a part of the future investigation. 

\item The field equations arising from the large N limit of all such systems are to be investigated (See, for example, \cite{soliton} and the references therein).

\item In the pseudo-euclidean metrics Chern-Simons gauge theory in the infrared region is
 associated with dissipative dynamics\cite{bla}. The connection of the balanced loss and gain systems considered in this work
with particular gauge theory will be very much interesting.

\item In the present work, we mainly concerned with the Hamiltonian formulation of a generic many-body 
systems with balanced loss and gain, their integrability and exact solvability at the classical level.
The quantization of this kind of systems will be a part of the future investigation.

\end{itemize}

\section{Acknowledgements}
This work is partly supported by a grant({\bf DST Ref. No. SR/S2/HEP-24/2012})
from Science \& Engineering Research Board(SERB), Department of Science
\& Technology(DST), Govt. of India. {\bf DS} acknowledges a research
fellowship from CSIR.

\section{Appendix A}

The choices of $V$ are ubiquitous for which the system governed by Eq.
(\ref{homo-eqn}) is exactly solvable. In this appendix, results of a
simple exactly solvable potentials, i.e solution for Henon-Heils system
\cite{henon} is presented. This system was first time introduced by 
 Henon and Heils in 1964 while examining the constants of motion in
galactic dynamics. The system is characterized by the potential
\be
V=\sum^{2}_{i=1}(-\omega^2_iz^2_i)- \frac{\alpha}{2} z^2_1z_2-\frac{1}{6}\beta
z_2^3.
\ee
\noindent There are three distinct regions in parameter-space for which exact
solutions exist\cite{ml}. For example, for the choice of the parameters
$\omega_1=\omega_2\equiv \omega_0, \beta =
-\alpha$, $z_i$ satisfy the equation:
\be
\ddot{z}_1 + \omega^2 {z}_1 + 2\alpha z_1 z_2 = 0,\
\ddot{z}_2 + \omega^2 {z}_2 + \alpha (z_1^2+ z_2^2)= 0, \
\omega^2=4(\omega_0^2 -\gamma^2).
\ee
\noindent These two equations are separable in the co-ordinate,
$u=z_1+z_2, v=z_1-z_2$ and describe two oscillators with quadratic interaction:
\be
\ddot{u} + \omega^2 u + \alpha u^2=0, \
\ddot{v} + \omega^2 v + \alpha v^2=0.
\ee
These Eqs. are exactly solvable with the solutions having the form\cite{amen}: 

\bea
u_i&=&A_i cn^2[\Omega_i t,k_i^2]+b_i, k_i^2=\frac{A_i^2 \alpha}{6 \Omega_i^2}, b_i=\frac{-4[\Omega_i^2(2k_i^2-1)+\omega^2]}{2\alpha},\ \ 0<k_i<1,
\nonumber\\
&&\Omega_i^4=\frac{\omega^4}{16(k_i^4-k_i^2+1)},\ \ \ i=1,2,\ \ u_1=u,u_2=v,
%v&=&A_1 cn[\Omega_1 t,k_1^2]+b_1, k_1^2=\frac{A_1^2 \alpha}{6 \Omega_1^2}, b_1=\frac{-4[\Omega_1^2(2k_1^2-1)+\omega]}{2\alpha},
%\Omega_1^4=\frac{\omega^4}{16(k_1^4-k_1^2+1)}
\eea
and 
\bea
z_1=\frac{1}{2}\left[u+v\right],\ \ \ \ z_2=\frac{1}{2}\left[u-v\right].
\eea
\bea
z_1^+=\gamma\left[A_1E(k_1,A_1)+A_2E(k_2,A_2)\right],\ \ \ \ z_2^+=\gamma\left[A_1E(k_2,A_2)-A_2E(k_2,A_2)\right],
\eea
where
\bea
E(k_i,A_i)=(t-\frac{t}{k_i^2})+\frac{EllipticalE[am[\Omega_i t, k_i^2],k^2](cn^2[\Omega_i t,k^2_i]+
\frac{1}{k_i^2}-1)}{\Omega_i dn[\Omega_i t,k_i^2]\sqrt{1-k_i^2sn^2[\Omega_i t, k_i^2]}}.
\eea
It should be mentioned here that the expression of $E(t)$ encounters singularity and therefore the 
solutions of $z_i^+$'s are not stable.

\section{Appendix B}

The Lagrangian of a N particle rotationally symmetric system in hyper-spherical coordinates
has the following form:
\bea
L=\frac{1}{2}\left[\dot{r}^2+r^2\dot{\theta}^2_1+r^2\left\{\sum^{N-1}_{i=2}\left(\dot{\theta_i^2}
\prod^{i-1}_{j=1}\sin^2\theta_j\right)\right\} \right]- V(r).
\eea
The corresponding Hamiltonian is:
\bea
H=\frac{1}{2}\left[p_r^2+\frac{p^2_{\theta_1}}{r^2}+\sum_{i=2}^{N-1}\frac{p^2_{\theta_i}}{r^2\prod_{j=1}^{i-1}\sin^2\theta_j}\right]
+V(r)
\eea
The Hamilton-Jacobi (HJ) characteristic function in this case may be written as:
\be
W=W_r(r)+\sum_{i=1}^{N-1}W_{\theta_{i}}(\theta_{i})+\alpha_{N-2}\theta_{N-1},
\label{W}
\ee
with $\theta_{N-1}$ being a cyclic coordinate.
The (HJ) equation takes the following form:
\bea
\left[\left(\frac{\partial W_r}{\partial r}\right)^2+\frac{1}{r^2}\left\{\left(\frac{\partial W_{\theta_{1}}}
{\partial \theta_{1}}\right)^2+\sum_{i=2}^{N-1}\frac{1}{\prod_{j=1}^{i-1}\sin^2\theta_j}\left(\frac{\partial 
W_{\theta_{i}}}{\partial \theta_{i}}\right)^2\right\}\right]
+2V(r)=2E.
\label{hj}
\eea
Now, the term in the curly bracket is only a function of the angular variables and must therefore be a 
constant ($=\alpha^2_r$).
From Eq. (\ref{hj}), we therefore have:
\bea
&&\left(\frac{\partial W_r}{\partial r}\right)^2+\frac{\alpha_r^2}{r^2}=2(E-V)
\label{wr}\\
&&\left(\frac{\partial W_{\theta_{1}}}
{\partial \theta_{1}}\right)^2+\sum_{i=2}^{N-1}\frac{1}{\prod_{j=1}^{i-1}\sin^2\theta_j}\left(\frac{\partial 
W_{\theta_{i}}}{\partial \theta_{i}}\right)^2=\alpha^2_r
\label{w_1}
\eea
The first Eq. (\ref{wr}) gives
\bea
W_r=\int \sqrt{2(E-V)-\frac{\alpha_r^2}{r^2}}dr
\label{1wr},
\eea
and the second Eq. (\ref{w_1}) may be written as:
\bea
\left(\frac{\partial W_{\theta_{1}}}
{\partial \theta_{1}}\right)^2+\frac{1}{\sin^2_{\theta_1}}\left\{\left(\frac{\partial W_{\theta_{2}}}
{\partial \theta_{2}}\right)^2+\sum_{i=3}^{N-1}\frac{1}{\prod_{j=2}^{i-1}\sin^2\theta_j}\left(\frac{\partial 
W_{\theta_{i}}}{\partial \theta_{i}}\right)^2\right\}=\alpha^2_r
\label{1w_1},
\eea
again the term in the curly bracket is devoid of $\theta_1$ and must be a constant $\alpha_{\theta_1}^2$.
Thus from Eq. (\ref{1w_1}), we have
\bea
W_{\theta_1}=\int \sqrt{\alpha^2_r-\frac{\alpha_{\theta_{1}^2}}{\sin^2{\theta_1}}}d{\theta_1}
\label{2w_1}.
\eea
By carrying on similar procedure we have in general:
\bea
W_{\theta_i}=\int \sqrt{\alpha^2_{i-1}-\frac{\alpha_{\theta_{i}^2}}{\sin^2{\theta_i}}}d{\theta_i}
\label{wthi}.
\eea
Thus if $\beta_r, \beta_{\theta_i} \forall i, i=1...N-1\ $ are the initial values of the radial and angular
coordinates respectively then from Eq. (\ref{W}) we have:

\bea
\beta_t+t&=&\frac{\partial W}{\partial E}=\int \frac{dr}{\sqrt{2(E-V)-\frac{\alpha^2_r}{r^2}}},
%\label{r}
\nonumber\\
\beta_{r}&=&\frac{\partial W}{\partial \alpha_r}=-\int \frac{\alpha_r  dr}{r^2\sqrt{2(E-V)-\frac{\alpha^2_r}{r^2}}}
+\int \frac{\alpha_r d\theta_1}{\sqrt{\alpha_r^2-\frac{\alpha^2_{\theta_{1}}}{\sin^2\theta_1}}},\nonumber\\
\beta_{\theta_{i}}&=&\frac{\partial W}{\partial \alpha_{\theta_i}}=-\int \frac{\alpha_{\theta_i} d\theta_i}
{\sin^2\theta_i\sqrt{\alpha_{\theta_{i-1}}^2-\frac{\alpha^2_{\theta_{i}}}{\sin^2\theta_i}}}
+\int \frac{\alpha_{\theta_i} d\theta_{i+1}}{\sqrt{\alpha_{\theta_{i}}^2-\frac{\alpha^2_{\theta_{i+1}}}{\sin^2\theta_{i+1}}}},\nonumber\\
\beta_{N-1}+\theta_{N-1}&=&\frac{\partial W}{\partial \alpha_{\theta_{N-1}}}=-\int \frac{\alpha_{\theta_{N-2}}\sin^2\theta_{N-2} d\theta_{N-2}}{\sin^2\theta_{N-2}\sqrt{\alpha_{N-3}^2-\frac{\alpha^2_{\theta_{N-2}}}{\sin^2\theta_{N-2}}}}.
\label{jhso}
\eea
Thus, we have $n+1$ integral Eqs. The first Eq. gives $r$ as function of time. The differential form of which may be written as:
\bea
\ddot{r}-\frac{\alpha_r^2}{r^3}+\frac{\partial V}{\partial r}=0,
\label{App}
\eea
which can be solved for the specific form of the potential $V$. It may be noted that in order to
map Eq. (\ref{App}) to that of considered in Sec. 3.1.2, one needs to replace $V$ by $-V$.
The second Eq. gives the relation between $r$ and $\theta_1$ and may be written as
\bea
\beta_{r}=\frac{\partial W}{\partial \alpha_r}=-\int \frac{\alpha_r  dr}{r^2\sqrt{2(E-V)-\frac{\alpha^2_r}{r^2}}}-
\sin^{-1}(\frac{\alpha_r}{\sqrt{\alpha_r^2-\alpha^2_{\theta_1}}}\cos\theta_1).
\eea
Rest of the Eqs. give relations between $\theta_i$ and $\theta_{i+1}$ $\forall i, i=2,.....N-2$. Thus, once the solutions
of the set of Eqs. in (\ref{jhso}) are known, the radial and angular variables can be expressed as a function of time. 

\section{Appendix C}

The Lax-pair for the system (\ref{h}) may be written as\cite{sata}:
\bea
L_{ij} &=& p_{z_ i}\delta{ij} + (1-\delta_{ij})\frac{i g}{\sqrt{2}(z_i-z_j)},\\
{\cal M}_{ij}&=&\frac{ig}{\sqrt{2}}\left[\delta_{ij}\sum_{l,(l\ne i)}^m\frac{1}{(z_i-z_l)^2}-(1-\delta_{ij})\frac{1}{(z_i-z_j)^2}\right]
\eea
The diagonal matrix $X(t)$ is defined by $X(t)=\delta_{ij} z_i(t)$. The Eq. (\ref{h}) may now be written
in the following matrix form:
\bea 
\dot{X}-[X,{\cal M}]=L,\ \ \ \ \dot{L}-[L,{\cal M}]=-\omega^2X
\eea
We define another matrix $Q(t)$ in the following fashion:
\be
Q(t)=U(t)X(t)U^{-1}(t),
\ee
where $U(t)$ is a unitary matrix , satisfying the relation:
\be
\dot{U}=U{\cal M},\ \ \ \ U(0)=1_{m\times m}.
\ee
With this constructions it is easy to evaluate that
\bea
\dot{Q}&=&U( \dot{X}- [X, {\cal M}])U^{-1} = ULU^{-1},\\
\ddot{Q}&=&U( \dot{L}-[L, {\cal M}])U^{-1} = -\omega^2Q
\eea
with the following solutions:
\bea
Q(t) = Q(0) \cos{(\omega t)} + \omega^{ -1} \dot{Q}(0) \sin{(\omega t)},
%\label{q}
\eea
where $Q(0)=X(0)$ and $ \dot{Q}(0)=L(0)$ are obtained from the initial values of $z_i(0)$ and $p_{z_ i}(0)$.

\section{Appendix D: $D_N$ type Calogero system with four-body interaction}

$D_N$ type Calogero system arises if we take the potential $V(z_i)$ as
\bea
V(z_i)=-\sum_{i}^m 2\omega_0^2 z^2_i-\sum_{\substack{i,j=1 \\ i < j}}^m\frac{g^2}{2(z_i-z_j)^2}-\sum_{\substack{i,j=1 \\ i < j}}^m\frac{g^2}{2(z_i+z_j)^2}
\eea
and the Eq. of motion (\ref{homo-eqn}) becomes:
\bea
&& \ddot{z}_i+\omega^2 {z}_i -\sum_{j,(j\ne i)}^m\left[\frac{g^2}{(z_i-z_j)^3}+\frac{g^2}{(z_i+z_j)^3}\right]= 0 
\label{h1},\\
&& z_i^+(t) = 2 \gamma \int z_i(t) dt + C_i, \ \ i=1, 2, \dots m.
\label{homo-eqn2}
\eea
The solution of Eq. (\ref{h1}) may be obtained
from the Lax-pair formulation\cite{sata,tyama}. The Lax-pair for the system (\ref{h1}) may be written as:
\bea
{\bf L}=
\bp 
L & S\\
-S & -L 
\ep, \  \
{\bf M}=
\bp 
{\cal M} & T\\
T & {\cal M}
\ep, \ \
{\bf X}=
\bp 
X & 0\\
0 & -X
\ep
\eea
where
\bea
L_{ij} &=& p_{z_ i}\delta{ij} + (1-\delta_{ij})\frac{i g}{\sqrt{2}(z_i-z_j)},\\
{\cal M}_{ij}&=&\frac{1}{\sqrt{2}}\left[\delta_{ij}\sum_{l,(l\ne i)}^m\left[\frac{ig}{(z_i-z_l)^2}+\frac{ig}{(z_i+z_l)^2}\right]-(1-\delta_{ij})\frac{ig}{(z_i-z_j)^2}\right],\\
S_{ij}&=&(1-\delta_{ij})\frac{i g}{\sqrt{2}(z_i+z_j)},\\
T_{ij}&=&-(1-\delta_{ij})\frac{ig}{\sqrt{2}(z_i+z_j)^2}
\eea
and the diagonal matrix $X(t)$ is defined by $X(t)=\delta_{ij} z_i(t)$. The Eq. (\ref{h1}) may now be written
in the following matrix form:
\bea 
\dot{{\bf X}}-[{\bf X},{\bf M}]={\bf L},\ \ \ \ \dot{{\bf L}}-[{\bf L},{\bf M}]=-\omega^2{\bf X}
\eea
We define another matrix $Q(t)$ in the following fashion:
\be
Q(t)=U(t){\bf X}(t)U^{-1}(t),
\ee
where $U(t)$ is a unitary matrix , satisfying the relation:
\be
\dot{U}=U{\bf M},\ \ \ \ U(0)=1_{m\times m}.
\ee
With this constructions it is easy to evaluate that
\bea
\dot{Q}&=&U(\dot{{\bf X}}-[{\bf X},{\bf M}])U^{-1} = U{\bf L}U^{-1},\\
\ddot{Q}&=&U(\dot{{\bf L}}-[{\bf L},{\bf M}])U^{-1} = -\omega^2Q
\eea
with the following solutions:
\bea
Q(t) = Q(0) \cos{(\omega t)} + \omega^{ -1} \dot{Q}(0) \sin{(\omega t)},
\label{q1}
\eea
where $Q(0)=X(0)$ and $ \dot{Q}(0)=L(0)$ are obtained from the initial values of $z_i(0)$ and $p_{z_ i}(0)$.
Thus, $z_i$'s are obtained from the eigen values of the matrix $Q(t)$ and $z_i^+$'s are obtained from the 
 eigen values of the matrix $R(t)$:
\be
R(t)= \frac{Q(0)}{\omega} \sin{(\omega t)} - \omega^{ -2} \dot{Q}(0) \cos{(\omega t)}
\label{nmsolu1}
\ee 
Once the eigen values of the matrices $Q(t)$ and $R(t)$ are known, $x_{2i-1}$ and $x_{2i}$ can
easily be constructed using Eq. (\ref{trans}).
After a coordinate transformation of the form (\ref{trans}), the Eq. (\ref{h1}) gives rise to the following 
Eqs. of motion:
\bea
&&\ddot{x}_{2l-1}+2\omega_0^2(x_{2l-1}-x_{2l})-2\gamma \dot{x}_{2l-1}\nonumber\\
&-&\sum_{\substack{i=1\\i\ne l}}^m
2g^2\left[\frac{1}{(x_{2i-1}-x_{2i}-x_{2l-1}+x_{2l})^3}+\frac{1}{(x_{2i-1}-x_{2i}+x_{2l-1}-x_{2l})^3}\right]=0,\nonumber\\
&& \ddot{x}_{2l}+2\gamma \dot{x}_{2l}-2\omega_0^2(x_{2l-1}-x_{2l})\nonumber\\
&+&\sum_{\substack{i=1\\i\ne l}}^m
2g^2\left[\frac{1}{(x_{2i-1}-x_{2i}-x_{2l-1}+x_{2l})^3}+\frac{1}{(x_{2i-1}-x_{2i}+x_{2l-1}-x_{2l})^3}\right]=0,\nonumber\\
&& l=1,..,m
\label{nm2}
\eea
with
\bea
V=-\sum_{i}^m \omega_0^2 (x_{2i-1}-x_{2i})^2-\sum_{\substack{i,j=1 \\ i < j}}^m
g^2\left[\frac{1}{(x_{2i-1}-x_{2i}-x_{2j-1}+x_{2j})^2}+\frac{1}{(x_{2i-1}-x_{2i}+x_{2j-1}-x_{2j})^2}\right].\nonumber
%\label{vfour1}
\eea

\end{document}